\documentclass[sigconf, nonacm, pdfa]{acmart}
\usepackage[a-2b]{pdfx}
\usepackage{graphicx}
\usepackage{subcaption}
\usepackage{adjustbox}
\usepackage{xspace}
\usepackage{tcolorbox}
\usepackage{ragged2e}
\usepackage{booktabs} 
\usepackage{multirow}  
\usepackage[ruled,vlined, linesnumbered]{algorithm2e}
\usepackage{xcolor}
\usepackage{soul}
\usepackage{tabularx}
\usepackage{makecell}
\SetKw{Continue}{continue}
\SetKw{Break}{break}
\usepackage{pifont}
\usepackage{balance}

\newcommand\vldbdoi{10.14778/3742728.3742757}
\newcommand\vldbpages{2681 - 2694}
\newcommand\vldbvolume{18}
\newcommand\vldbissue{8}
\newcommand\vldbyear{2025}
\newcommand\vldbauthors{\authors}
\newcommand\vldbtitle{\shorttitle} 
\newcommand\vldbavailabilityurl{https://github.com/VIDA-NYU/magneto-matcher}
\newcommand\vldbpagestyle{empty} 

\theoremstyle{definition}

\theoremstyle{definition}
\newtheorem{definition}{Definition}[section]

\newcommand{\hide}[1]{}
\newcommand{\myparagraph}[1]{\vspace{0.25em}\noindent \textbf{#1.}}

\newcommand{\attr}[1]{\texttt{#1}}

\newenvironment{myitemize}%
{\begin{list}{$\bullet$}{%
			\setlength{\labelsep}{2pt}\setlength{\leftmargin}{5pt}%
			\setlength{\labelwidth}{0pt}%
			\setlength{\listparindent}{0pt}}}
	{\end{list}}

\def\withnotes{1} 

\definecolor{HighlightColor}{rgb}{0.05,0.05,0.70}
\newcommand{\revision}[1]{{\color{HighlightColor}#1}}
\renewcommand{\revision}[1]{#1}

\ifnum\withnotes=1    
    \definecolor{Maroon}{rgb}{0.62, 0.0, 0.09}
    \definecolor{Emerald}{rgb}{.07, .74, .62}
    
    \newcommand{\ep}[1]{\textcolor{teal}
    {\textbf{Pena:} #1}}
    \newcommand{\as}[1]{\textcolor{violet}{\textbf{Aécio:} #1}}
    \newcommand{\jf}[1]{\textcolor{Emerald}{\textbf{Juliana:} #1}}
    \newcommand{\yurong}[1]{\textcolor{orange}{\textbf{Yurong}: #1}}
    \newcommand{\eden}[1]{\textcolor{teal}{\textbf{Eden}: #1}}
    \newcommand{\roque}[1]{\textcolor{violet}{\textbf{Roque}: #1}}
    \newcommand{\draft}[1]{\textcolor{red}{#1}}
\else
    \newcommand{\ep}[1]{}
    \newcommand{\as}[1]{}
    \newcommand{\jf}[1]{}
    \newcommand{\eden}[1]{}
    \newcommand{\draft}[1]{}
    \newcommand{\roque}[1]{}
    \newcommand{\yurong}[1]{}
\fi

\definecolor{RoyalBlue}{RGB}{65,105,225}
\definecolor{blue(ncs)}{rgb}{0.0, 0.53, 0.74}
\definecolor{bluegray}{rgb}{0.4, 0.6, 0.8}
\definecolor{cobalt}{rgb}{0.0, 0.28, 0.67}
\definecolor{mediumelectricblue}{rgb}{0.01, 0.31, 0.59}
\definecolor{bleudefrance}{rgb}{0.19, 0.55, 0.91}
\definecolor{maroon(x11)}{rgb}{0.69, 0.19, 0.38}
\definecolor{mangotango}{rgb}{1.0, 0.51, 0.26}
  
\newcommand{\ignore}[1]{\leavevmode\unskip}

\newcommand{\alg}{\texttt{Magneto}\xspace}

\newcommand{\algzsbp}{\texttt{Magneto-zs-bp}\xspace}
\newcommand{\algzsgpt}{\texttt{Magneto-zs-llm}\xspace}

\newcommand{\algftbp}{\texttt{Magneto-ft-bp}\xspace}
\newcommand{\algftgpt}{\texttt{Magneto-ft-llm}\xspace}

\newcommand{\ourapproach}{\texttt{Magneto}\xspace}
\newcommand{\ft}{\texttt{ft}\xspace}
\newcommand{\gpt}{\texttt{llm}\xspace}
\newcommand{\bp}{\texttt{bp}\xspace}

\newcommand{\llmaugfull}{LLM-based augmentation\xspace}
\newcommand{\llmaug}{\texttt{llm-aug}}
\newcommand{\structaugfull}{structure-based augmentation\xspace}
\newcommand{\structaug}{\texttt{struct-aug}}

\newcommand{\GDC}{GDC\xspace}
\newcommand{\GDCBenchmark}{\GDC benchmark\xspace}

\newcommand{\coma}{\texttt{COMA}\xspace}
\newcommand{\hlc}[2][yellow]{{%
    \colorlet{foo}{#1}%
    \sethlcolor{foo}\hl{#2}}%
}

\newcommand{\smabrev}{SM\xspace}
\renewcommand{\smabrev}{schema matching\xspace}
\newcommand{\Smabrev}{Schema matching\xspace}

\newcommand{\sm}{schema matching\xspace}

\begin{document}

\title{\alg: Combining Small and Large Language Models for Schema Matching}


\author{Yurong Liu}
\affiliation{%
  \institution{New York University}
  \city{}
	\state{}
}
\email{yurong.liu@nyu.edu}
\authornote{Equal contribution.}

\author{Eduardo H. M. Pena}
\affiliation{%
	\institution{Federal University of Technology~Paran\'{a}}
	\city{}
	\state{}
}
\email{eduardopena@utfpr.edu.br}
\authornotemark[1]
\authornote{Work done as a visiting researcher at New York University.}

\author{A\'ecio Santos }
\affiliation{%
  \institution{New York University}
}
\email{aecio.santos@nyu.edu}

\author{Eden Wu}
\affiliation{%
  \institution{New York University}
}
\email{eden.wu@nyu.edu}

\author{Juliana Freire}
\affiliation{%
  \institution{New York University}
}
\email{juliana.freire@nyu.edu}

\begin{abstract}

Recent advances in language models (LMs) open new opportunities for schema matching (SM).  Recent approaches have shown their potential and key limitations: while small LMs (SLMs) require costly, difficult-to-obtain training data, large LMs (LLMs) demand significant computational resources and face context window constraints. We present \ourapproach, a cost-effective and accurate solution for SM that combines the advantages of SLMs and LLMs to address their limitations. By structuring the SM pipeline in two phases, retrieval and reranking, \alg can use computationally efficient SLM-based strategies to derive candidate matches which can then be reranked by LLMs, thus making it possible to reduce runtime while improving matching accuracy. We propose (1) a self-supervised approach to fine-tune SLMs which uses LLMs to generate syntactically diverse training data, and (2) prompting strategies that are effective for reranking. We also introduce a new benchmark, developed in collaboration with domain experts, which includes real biomedical datasets and presents new challenges for SM methods. Through a detailed experimental evaluation, using both our new and existing benchmarks, we show that \alg is scalable and attains high accuracy for datasets from different domains.

%
\end{abstract}

\maketitle


\pagestyle{\vldbpagestyle}
\begingroup\small\noindent\raggedright\textbf{PVLDB Reference Format:}\\
\vldbauthors. \vldbtitle. PVLDB, \vldbvolume(\vldbissue): \vldbpages, \vldbyear.\\
\href{https://doi.org/\vldbdoi}{doi:\vldbdoi}
\endgroup
\begingroup
\renewcommand\thefootnote{}\footnote{\noindent
This work is licensed under the Creative Commons BY-NC-ND 4.0 International License. Visit \url{https://creativecommons.org/licenses/by-nc-nd/4.0/} to view a copy of this license. For any use beyond those covered by this license, obtain permission by emailing \href{mailto:info@vldb.org}{info@vldb.org}. Copyright is held by the owner/author(s). Publication rights licensed to the VLDB Endowment. \\
\raggedright Proceedings of the VLDB Endowment, Vol. \vldbvolume, No. \vldbissue\ %
ISSN 2150-8097. \\
\href{https://doi.org/\vldbdoi}{doi:\vldbdoi} \\
}\addtocounter{footnote}{-1}\endgroup

\ifdefempty{\vldbavailabilityurl}{}{
\vspace{.3cm}
\begingroup\small\noindent\raggedright\textbf{PVLDB Artifact Availability:}\\
The source code, data, and/or other artifacts have been made available at \url{\vldbavailabilityurl}.
\endgroup
}


\section{Introduction }
\label{sec:intro}


The rapid increase in the volume of structured data--   from data published in scientific articles~\cite{nature,science} and repositories~\cite{gdc,pdc,ukbiobank,zenodo} to open government portals~\cite{nycopendata,datagov} --  creates new opportunities to answer important questions through analytics and predictive modeling. But often, data from multiple sources must be integrated to answer these questions. Consider the following example.

\begin{example}{} 
\label{example:p-analysis}
In the context of the National Cancer Institute’s Clinical Proteomic Tumor Analysis Consortium (CPTAC)~\cite{cptac}, \citet{li2023proteogenomic} carried out a comprehensive proteogenomic analysis of cancer tumors. They collected data 
from ten studies (published as supplementary material in research papers) that cover multiple patient cohorts and cancer types. To facilitate their analysis, they mapped each dataset into the GDC standard,  a data model set by the National Cancer Institute's Genomic Data Commons (GDC) for cancer genomic data~\cite{heath2021nci,gdc}.
Even though the studies had been carried out by members of the CPTAC effort, the datasets containing patient case and sample data used  different representations for variable names and values.  
Integrating these data required a substantial effort to match variables from each dataset source schemata to the target GDC format, which encompasses over 700 attributes. 
\qed
\end{example}
\vspace{-0.5em}

\begin{figure}[t]
    \centering
    \includegraphics[width=.75\columnwidth]{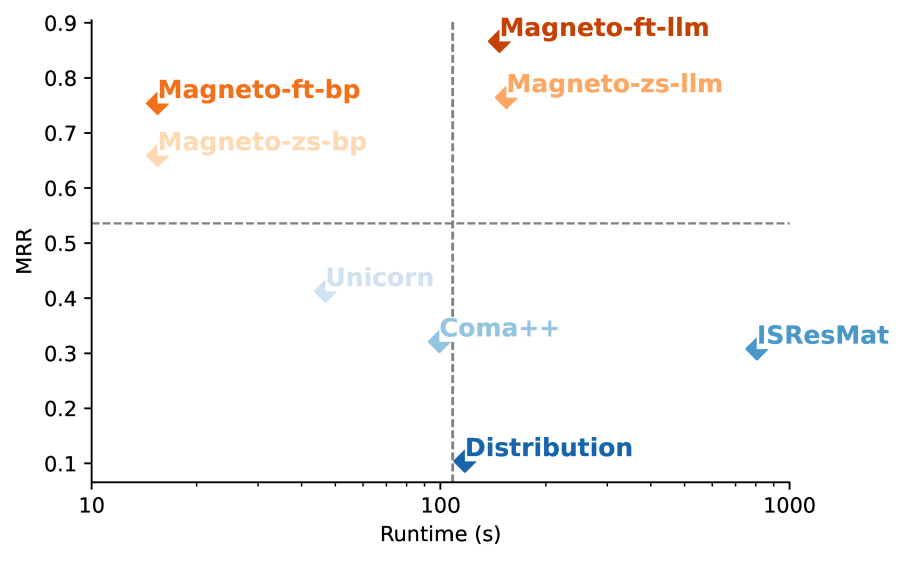}
    \vspace{-.2cm}
    \caption{
    Trade-off between runtime and accuracy (using MRR) for different schema matchers in the task of Example~\ref{example:p-analysis}. \alg variants (orange) outperform traditional methods (blue) in MRR and accuracy-runtime trade-off.}
   \vspace{-.4cm}
    \label{fig:motivation_mrr}
\end{figure}

Even though the \sm 
problem has been extensively studied~\cite{doan2012integrationBook,cafarellaIntegrationonweb2009,miller2018opendataintegration,valentine2021}, 
%
matching still requires a time-consuming, manual curation process for complex tasks, which like the one described above,
involves ambiguity and heterogeneity in the representation of attributes and values (see Table~\ref{table:match-examples}). 
\Smabrev approaches that rely on attribute names, data types and values
for similarity assessments are likely to fail for such matches.
As a point of reference, we assessed the effectiveness of state-of-the-art strategies for the biomedical datasets from Example~\ref{example:p-analysis}. As shown in Figure~\ref{fig:motivation_mrr}, they perform poorly: the best technique achieves at most 0.45 of mean reciprocal rank (MRR) and incurs high computational costs.
We discuss these results in detail in Section~\ref{sec:experiments}.

Determining correspondences between columns may require knowledge beyond the schema and contents of a table.
Table~\ref{table:match-examples} shows some possible matches for the proteogenomic analysis.
Sometimes, selecting the correct match is difficult even for subject matter experts.  For example, for the attribute \attr{patient\_age} from one of the datasets, there are at least three plausible matches in GDC: \attr{age\_at\_diagnosis}, \attr{days\_to\_birth}, and \attr{age\_at\_index}. Without additional context, it is difficult to determine the correct match. 


\myparagraph{Schema Matching and Language Models}
Renewed interest in data integration has emerged due to the capabilities of language models~\cite{archetypeVLDB2024,llmWrangle2022}. For \smabrev, promising approaches have been proposed~\cite{SEMPROPicde2018, cappuzzo2020creating,inSituSchMatcICDE2024,tu2023unicorn}. A key challenge in \smabrev is estimating the similarity between two columns. Pre-trained Language Models (PLMs), referred to as SLMs in this paper to distinguish them from large language models (LLMs), create column representations (or embeddings) enriched with semantic information. The similarity between two embeddings serve as a proxy for column-matching scores~\cite{SEMPROPicde2018, cappuzzo2020creating}. SLMs have also been fine-tuned for schema matching~\cite{inSituSchMatcICDE2024} and general matching tasks~\cite{tu2023unicorn}.

\begin{table}[t]
\centering
\caption{
Examples of candidate matches between source tables and the GDC that highlight schema heterogeneity. 
}
\label{table:match-examples}
\footnotesize
\renewcommand{\arraystretch}{1.5} 
\setlength{\tabcolsep}{6pt} 
\begin{tabular}{>{\raggedright\arraybackslash}m{4.1cm}>{\raggedright\arraybackslash}m{3.5cm}}
\toprule 
\textbf{Source Column and Values}  & \textbf{Target Candidate Columns} \\ 
\midrule
\hlc[cyan!0]{\textbf{Histologic\_Grade\_FIGO}:}  FIGO grade 1,  FIGO grade 2,  FIGO grade 3 & 
\hlc[cyan!0]{\textbf{tumor\_grade}:} G1, G2, G3 \newline 
\hlc[cyan!0]{\textbf{who\_nte\_grade}:} G1, G2, G3  \newline
\hlc[cyan!0]{\textbf{adverse\_event\_grade}:} Grade 1, Grade 2, Grade 3
\\ \hline
\hlc[cyan!0]{\textbf{Path\_Stage\_Primary\_Tumor-pT}:}  pT1a (FIGO IA),  pT2 (FIGO II),  pT3b (FIGO IIIB) &
\hlc[cyan!0]{\textbf{uicc\_pathologic\_t}:} T1a, T2, T3b  \newline
\hlc[cyan!0]{\textbf{ajcc\_pathologic\_t}:} T1a, T2, T3b
\\ \hline
\hlc[cyan!0]{\textbf{Age}:}  65,  72,  49 &
\hlc[cyan!0]{\textbf{age\_at\_diagnosis}:} n/a \newline 
\hlc[cyan!0]{\textbf{days\_to\_birth}:} n/a  \newline 
 \hlc[cyan!0]{\textbf{age\_at\_index}:} n/a 
\\ 
\bottomrule
\end{tabular}
\end{table}

LLMs, in contrast, are trained using large, generic data corpora and thus contain knowledge that can assist in obtaining additional semantics necessary to identify matches. Prompting strategies combined with fine-tuned models have been shown effective to improve table understanding and help in integration tasks~\cite{tableGPT2024}.

While prior work has shown the usefulness of SLMs and LLMs for \smabrev, they also present significant practical challenges:
\begin{myitemize}
    \item \emph{Challenge 1}: Fine-tuning SLMs can lead to significant performance improvements, but this requires the availability of manually curated training data, which can be expensive to create.
    \item \emph{Challenge 2}: \revision{LLMs usually do not need fine-tuning, but face constraints due to fixed context windows, requiring truncation of large prompts and potential loss crucial information. While some models offer larger windows, cost scales with input and output size. Furthermore, accuracy can decline with long prompts and API calls incur high latency, especially for large inputs~\cite{li-long-prompt@corr2024}.}
\end{myitemize}

\begin{figure}[t]
    \centering
    \includegraphics[width=0.7\linewidth]{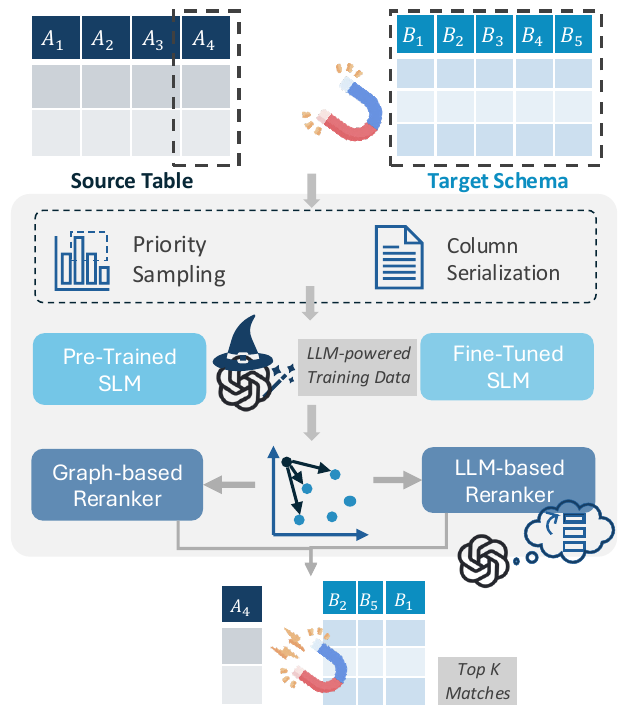}
    \vspace{-.2cm}
    \caption{
    \alg takes source and target tables and identifies matches in two phases: an SLM retrieves and ranks candidates, then an LLM assesses and reranks them. This enables efficient matching and cross-domain generalization, with customizable strategies based on method combinations.
    }
    \vspace{-.5cm}
    \label{fig:framework}
\end{figure}

\myparagraph{Our Approach}
%
We introduce \ourapproach, a framework that combines SLMs and LLMs to derive cost-effective and general \smabrev solutions. As illustrated in Figure~\ref{fig:framework}, \ourapproach is structured in two phases: \emph{candidate retrieval}  selects a subset of the possible matches; and reranker, which
ranks the candidates to make it easier for users to examine and select matches.
%
To address \emph{Challenge 1}, \alg leverages LLMs to automate SLMs fine-tuning. Instead of relying on manually created training data and structure-based augmentation (e.g., row shuffling and sampling \cite{starmie2023,inSituSchMatcICDE2024}), we use LLMs to derive data. Using LLMs, we can add syntactic diversity and capture different representations for column names and their values. 
\alg addresses \emph{Challenge 2} by using cheaper SLM-based methods for finding candidates thereby reducing the number of matches that need to be checked by more costly LLM-based rerankers. 
To ensure that all necessary details are included in the prompt while staying within the context window limit, we need to create representative samples of the 
columns in candidate matches to be ranked by the LLM. 
In addition, since LLMs are designed for textual data,  we must serialize the column content.
Selecting the right serialization strategy is still an open research problem that has attracted substantial attention~\cite{fang2024large,llm-table@sigir2024}. 
We explore different alternatives for both sampling and serialization.  


Our goal with \alg is to achieve high accuracy for \smabrev at a low cost.
Figure \ref{fig:motivation_mrr} illustrates the trade-offs between runtime and accuracy for existing \smabrev algorithms and different variants of \alg.
Note that traditional methods tend to cluster at relatively high accuracy with long runtimes or relatively low runtimes with reduced accuracy. In contrast, \alg variants strike a balance across both dimensions and have the advantage of requiring no training data.


%
We conduct an extensive experimental study of \alg on various datasets from two benchmarks. The first is a new benchmark, that we developed in collaboration with domain experts, which includes real biomedical datasets. 
Our experiments show that many existing \smabrev solutions struggle with this benchmark, particularly given the complexity and syntactic variability present in its datasets.
In contrast, \alg demonstrates superior accuracy and generally faster runtimes.
We also evaluate \alg on the Valentine 
benchmark~\cite{valentine2021} and observe that it also performs well for the Valentine datasets which cover a variety of domains.

\myparagraph{Contributions} 
Our main contributions are summarized as follows:
\begin{myitemize}
\item We introduce the \alg framework that effectively combines SLMs and LLMs and allows the creation of \smabrev strategies that attain a balance between accuracy and runtime for diverse data and tasks (Section~\ref{sec:background}). 
\item We propose an LLM-powered method to generate training data for fine-tuning SLMs for \smabrev tasks and a contrastive learning pipeline using triplet loss and online triplet mining to enhance column embedding distinction (Section~\ref{sec:match_retrieval}).

\item Unlike previous approaches that use LLMs for \smabrev, 
we show that reranking candidate matches with LLMs, derived from less costly methods, achieves high accuracy at a lower cost. (Section~\ref{sec:match_selector}).
\item In collaboration with biomedical researchers, we created a new benchmark that represents a real schema matching effort~\cite{li2023proteogenomic},  encompassing characteristics not present in existing benchmarks and introducing new challenges for SM methods (Section~\ref{sec:benchmark}).
\item We perform an extensive experimental evaluation of different strategies derived using our framework, comparing them against SM methods over datasets that cover a wide range of domains. We also carry out ablation studies to assess the effectiveness of different choices for column sampling and serialization (Section~\ref{sec:experiments}).
\end{myitemize}


\section{Motivation and Solution Overview}
\label{sec:background}

\myparagraph{Biomedical Data Integration: Opportunities and Challenges} 
Due to substantial investments made in infrastructure to share biomedical data~\cite{nih_data_sharing_policy_2020}, many datasets are now stored in repositories with related research articles~\cite{pdc,gdc}. Pooling these data can advance research across various diseases and populations, but current technologies are limited in their ability to integrate these data. Our research was motivated by this challenge as part of a collaboration with biomedical researchers in the context of the ARPA-H Biomedical Data Fabric (BDF) program~\cite{arpahbdf}, which aims to develop usable data integration methods and tools.

\hide{In a recent project, we collaborated with medical researchers who are used to integrating biomedical datasets.  
They regularly integrate diverse datasets from patient registries, clinical trials, and genomic databases to develop and test hypotheses in complex fields, such as cancer research~\cite{fenyoNature2016}.
They depend on effective integration to improve data quality and volume since these are critical to reducing the biases and inconsistencies in their analysis.
Among their typical data sources are study results from researchers on a similar topic, which present significant variations in semantics, acquisition protocols, data formats, and scales. Such data sources present unique challenges for schema-matching, which is usually the first step in their integration pipeline.}

Our interviews with biomedical experts revealed that current data integration practices rely heavily on manual processes and dataset-specific scripts for \smabrev~\cite{fenyoNature2016, li2023proteogenomic, harmonizationPrime2024}. As illustrated in Example~\ref{example:p-analysis}, this approach is both error-prone and difficult to reproduce, creating significant barriers to  biomedical research.

Based on these findings, we identified two critical requirements that guided the design of Magneto. First, researchers need approaches that generalize across the diverse landscape of biomedical data—spanning different data types (genomics, proteomics, electronic health records), disease categories (various cancers, autoimmune and rare diseases), and data sources (publications, hospitals, data commons). Second, given the complexity of biomedical schemas, these approaches must facilitate curation, as even subject matter experts often struggle to determine correct matches. 

\hide{
These insights shaped our critical requirements for a practical \yurong{an accurate and efficient} schema matching solution: (1) facilitating the life of experts by showing potentially good quality matches in a ranked list; (2) delivering fast results to support an interactive workflow; and (3) achieving strong zero-shot performance.}

\subsection{Definitions and Evaluation Metrics}
\label{subsec:definitions}
Before describing our approach, we introduce the notation, schema matching definition, and evaluation metrics used in this paper.


\begin{definition}{\em{\bf (Schema Matching)}}
Let $\mathbf{S} (A_1, \dots, A_n)$ be a source table and $\mathbf{T}(B_1, \dots, B_m)$ be a target table, where $A_i \in \mathbf{S}$ and $B_j \in \mathbf{T}$ are columns that define the schemata. Each column $A \in \mathbf{S} \cup \mathbf{T}$ has an associated \emph{domain}, denoted $\mathcal{D}(A)$, representing the set of possible values that the column can take; note that $\mathcal{D}(A)$ may be empty. 
Schema matching focuses on aligning the table schemata by establishing correspondences between columns representing the same real-world concept or entity. A matching algorithm (or matcher) aims to identify pairs $(A_i, B_j)$ that represent the same (or semantically equivalent) column based on various factors, such as their domains and names.
Thus, a matcher $\mathcal{M}$ can be seen as a function that generates a schema mapping $M \subseteq \mathbf{S} \times \mathbf{T}$, where each element $(A, B) \in M$ represents a correspondence between a source column $A$ and a target column $B$, 
where $\mathcal{D}(A) \approx \mathcal{D}(B)$, meaning that the domains of columns $A$ and $B$ are related or overlap. 
\qed
\end{definition}

%
Matching algorithms often associate a score with each match they derive. These scores are used to generate ranked lists containing the derived matches, which help users explore matches by prioritizing the highest-scoring candidates. These lists provide a global ranking of best matches $(A_i, B_j)$ among all pairs of possible matches of a given pair of tables $\textbf{S}$ and $\textbf{T}$ or a per-column ranking of the best matches $B_j$ for a given source attribute $A_i \in \textbf{S}$.
Thus, a common approach to evaluate schema matching methods is to assess their ability to produce high-quality ranked lists of matches~\cite{valentine2021, inSituSchMatcICDE2024}. 
\revision{
We use two evaluation metrics: Mean Reciprocal Rank (MRR) and Recall at Ground-Truth Size (Recall@GT), as detailed next.
}


\begin{definition}{\em{\bf (Mean Reciprocal Rank)}}
Let $matches[A]$ denote the ranked list of matches produced by a schema-matching algorithm for the source column $A \in \textbf{S}$, and $r_A$ 
be the position of the first correct target column $B$ within the ordered list $matches[A]$.
The reciprocal rank (RR) of an individual column $A$ is the multiplicative inverse of the rank, i.e., $\frac{1}{r_A}$. 
The \textit{mean reciprocal rank} (MRR) for a table~$\textbf{S}$ is the average RR over the subset of columns $\textbf{S}'$ that contain a correct match in the ground truth:
\begin{equation}
   \text{MRR} = \frac{1}{|\textbf{S}'|} \sum_{A \in \mathbf{S}'} \frac{1}{r_A}.
\label{eq:mrr}
\end{equation}
%
Intuitively, \revision{MRR} measures how long it takes to find the first relevant match when examining the ranked list of matches. 
\qed
\end{definition}
\revision{
MRR is a standard evaluation metric for ranked lists in search engines and question-answering systems~\cite{schutze2008introduction, voorhees1999trec}. For \smabrev, high-quality results correspond to ranking the most relevant match for each source column as highly as possible. A high MRR score therefore indicates that users can more easily identify correct matches when evaluating candidate matches for a given column.}


%

We also use the recall at ground truth size (Recall@GT), a standard measure used in recent schema matching literature~\cite{valentine2021}.
Unlike MRR, which evaluates rankings per source column, Recall@GT operates on a global ranking that merges all candidate matches across columns into a single list.
%
\begin{definition}{\em{\bf (Recall@GT)}}
Let $matches$ denote the global ranked list containing all matches produced by a matcher that considers all pairs of possible matches between columns from $\textbf{S}$ and $\textbf{T}$, and let $\mathcal{M}$ denote a set containing only the top-$k$ best results in $matches$. 
Moreover, let $\mathcal{G}$ be the set of ground truth matching pairs $(A, B)$. 
Recall@GT measures the fraction of relevant matches in the ground truth that also appears in $\mathcal{M}$,
where $k$ is given by the size of the ground truth,  $k=|\mathcal{G}|$. More formally,
\begin{equation}
    \text{Recall@GT} = \frac{ 
    | 
    \mathcal{G} \cap \mathcal{M} 
    |
    }{k}.
\label{eq:recall@gt}
\end{equation}
Intuitively, it measures how well the matcher can place all correct matches of a table $\textbf{S}$ at the top of the global ranked list.
\qed
\end{definition}

\subsection{\alg: Overview}
\label{subsec:overview}


%
\alg first applies a cheaper approach to retrieve and filter candidates so that a more sophisticated method can accurately identify the correct matches from a smaller candidate set. 
%
Figure~\ref{fig:framework} shows one variant of \alg that uses a fine-tuned language model as the retriever and a large language model as the reranker.
%

\myparagraph{Candidate Retriever}
The \emph{candidate retriever} leverages an SLM to generate a ranked list of potential matches from the target table for each input column. 
It uses column embeddings to estimate column pair similarities~\cite{cappuzzo2020creating,starmie2023}. SLMs are a good choice for this step given their ability to capture semantic similarity and efficiency. 
\hide{and explores new variations of this approach. Our retriever achieves high accuracy with SLMs even without fine-tuning, given our proposed techniques and methods of utilization, which we refer to as the base \alg.}
\hide{A key component of \alg is the \emph{candidate retriever},
which returns a ranked list of columns in the target table as potential matches for each input column. We build on recent trends and use column embeddings to estimate column pair similarities~\cite{cappuzzo2020creating,starmie2023}, exploring new variations of this approach. Our retriever achieves strong accuracy even when using standard SLMs. 
Still, this component focuses on computational efficiency, and in some cases, the highest accuracy requirements may not be fully met.}
General-purpose pre-trained SLMs such as BERT~\cite{devlin2018bert}, RoBERTa~\cite{liu2019roberta}, and MPNet~\cite{song2020mpnet}, may struggle with syntactic differences and lack contextual knowledge for domains absent in their training data. 
To perform complex tasks effectively, these models need to be fine-tuned. We propose to leverage LLMs to automatically  generate the data needed to fine tune SLMs. We show that LLMs can help derive high-quality training data that reflect instances of variability for semantically similar columns that arise in real data. This approach, described in Section~\ref{sec:match_retrieval}, improves the robustness of the SLM-based retriever, making it possible for it to handle complex matches without requiring human-labeled training data. 
We refer to \alg configurations that use this approach using the label~\ft.
Note that \alg-\ft invokes LLMs only during the offline training phase for a given domain. Once trained, the fine-tuned SLM can perform multiple inferences efficiently without making calls to LLMs.

\hide{We propose fine-tuning strategies to improve the retriever's accuracy. 
We model this step by having schema matching as the optimization goal so that the fine-tuning can be effective.
In practice, fine-tuning can be challenging due to the sensitivity of SLMs to training data aspects. We propose incorporating large language models (LLMs) for automated data generation to serve fine-tuning SLMs. 
We generate high-quality training data that reflect complex cases of semantically similar columns. Our goal is to improve the robustness of our underlying model without requiring users to provide us with any prior training data.}

\myparagraph{Match Reranker}
While fine-tuning improves SLM performance, it is not sufficient to handle schema matching tasks involving domains and heterogeneity unseen during the fine tuning. To improve generalizability, \alg uses LLMs as \textit{rerankers} to refine the candidates identified by the SLM-based retriever. This approach, which we refer to as \alg-\gpt, enhances accuracy by leveraging carefully designed prompts and techniques that enable the LLM to judiciously assess matches and discern subtle semantic nuances, which are challenging for the SLM to detect independently. As discussed in Section~\ref{sec:match_selector}, \alg-\gpt also reduces
LLM costs.

\hide{The number of potential match candidates for each column in the source table can become overwhelming. This creates two key challenges: managing the combinatorial explosion of possible match combinations and establishing reliable criteria for selecting the most relevant matches.
The match reranker component addresses these by reordering all candidate matches to promote the ``best'' matches to the top of the list presented to the user.

We propose implementing this reranking using LLMs, building on recent advances in column type annotation~\cite{archetypeVLDB2024}. Our approach achieves higher accuracy and low cost by using carefully crafted prompts to guide the LLM in capturing subtle semantic relationships that may be difficult for the retriever alone to distinguish. }

\myparagraph{Varying Retrievers and Rerankers} The architecture of \alg allows the combination of different alternatives for retrievers and rerankers.
We can combine traditional matching techniques with the embedding strategies for the candidate retriever. For example, when columns in a match have the same name, \revision{modulo} minor variations in case and punctuation, we assign a perfect similarity score of 1.0.
During our experiments, this simple technique consistently improved overall accuracy, albeit slightly, in most cases.

\hide{
\jf{Eduardo: please check this}
Although exceptions do exist, this simple strategy
%
reduces the number of calls to LLMs by not sending \emph{easy matches} to be assessed, thus reducing cost and latency.
}


We also implemented \emph{bipartite-graph reranker}
as an algorithmic alternative that adapts the filtering technique from Melnik et al.~\cite{melnik2002similarity}. This approach, which we refer as \alg-\bp,  is particularly  
\revision{
suitable for scenarios where 
LLMs are unavailable during inference}
or where strict runtime constraints must be met.
The algorithm combines all ranked match candidates across source columns into a single global list and transforms it into an undirected bipartite graph.
There are two (disjoint) node sets consisting of the columns in the source and target tables (respectively).
The graph contains two disjoint node sets representing the source and target table columns, respectively, with potential matches represented as weighted edges between nodes. Edge weights correspond to match confidence scores.
%
To identify the \emph{best} match for the set of attributes in the source column, it uses the algorithm from~\cite{linearAssigment2016}, which solves the assignment problem in polynomial time and scales to large graphs.
The final ranking prioritizes matches selected by the assignment algorithm at the top of the list, while unselected matches are placed at the bottom in their original relative order.
\section{Using LLMs to Fine-Tune SLMs}
\label{sec:match_retrieval}

Pre-trained small language models (SLMs) have been used to help with schema-matching related tasks by encoding semantic information from column names and values into dense vector representations--\textit{embeddings}~\cite{tu2023unicorn, schemaMatchPreTrainedICDE2023, starmie2023, cong2023pylon, deepJoin2023, inSituSchMatcICDE2024}. 
To identify matches, embeddings of source and target columns can be compared using \textit{cosine similarity}: high similarity scores indicate a higher likelihood that the columns match. SLMs work well for general natural language tasks, but their ability to interpret tabular data is limited. These models must often be fine-tuned to handle tasks that involve tables.  
However, fine-tuning approaches require large amounts of labeled data for training~\cite{tu2023unicorn,deepJoin2023}.
Unfortunately, this is impractical in many scenarios such as the integration of biomedical data (Example~\ref{example:p-analysis}), for which training data is hard to obtain.

In the absence of training data, it is possible to apply augmentation techniques to automatically generate variations of data to be used as positive examples. For example, given a column, different versions can be generated by shuffling rows, sampling values, and applying perturbations to values~\cite{starmie2023,deepJoin2023}. However, these variations may not fully capture the heterogeneity found in real data. 
We introduce a new method that leverages LLMs to generate training data and present a pipeline to fine-tune SLMs for schema matching (Section~\ref{sec:modeltraining}).
Another key consideration is how to represent tables and their columns, which we discuss in Section~\ref{sec:serialization_sampling}.


\subsection{Value Sampling and Column Serialization}
\label{sec:serialization_sampling}

\myparagraph{Priority Sampling for Column Values}
We generate embeddings to retrieve candidate matches by including a sample of column values in the column representation.
%
Sampling strategies have been adopted for data management tasks, including  column type annotation~\cite{archetypeVLDB2024} and table union search~\cite{starmie2023}. A common approach is to use weighted sampling and assign higher weights to more frequent values. We adopt this approach and incorporate coordination into the sampling process with \textit{priority sampling}~\cite{daliri2023sampling}. 
For inner product sketching, priority sampling maximizes the likelihood of selecting corresponding values across vectors by emphasizing elements with larger magnitudes.

\revision{In our setting, priority sampling is adapted to optimize the selection of column values. %
This approach not only prioritizes frequent values, which are statistically more representative of the column domain, but it also increases the likelihood of identifying shared values across different columns that act as inter-column anchors. These anchors enhance similarity detection in SLMs, which are sensitive to token co-occurrence patterns learned during pre-training.
}
Specifically, we use a random seed $s$ to select a uniformly random hash function $h: \{1, \dots, N\} \to [0, 1]$, where $N$ represents the maximum number of unique values across all columns. For each value $v_i$, we compute a rank:
$$
R_i = \frac{\text{freq}(v_i)}{h(v_i)},
$$
where $\text{freq}(v_i)$ denotes the frequency of $v_i$. Then, we select the sample's first $m$ values with the largest priorities $R_i$. 
\myparagraph{Column Serialization}
Since SLMs interpret inputs as sequences of tokens (i.e., regular text), we must transform each column into a token sequence that the model can process. Recent research has explored various approaches to serializing columns into a textual format~\cite{taclPaolo2023}. Here, we explore the impact of different approaches to column serialization for schema matching.

Given column $A$ with column type $type(A)$ and sample values $v_1, v_2, \dots, v_m$, we consider the following serialization approaches:
\begin{small}
\begin{align*}
     \mathcal{S}_{\texttt{default}}(A) &= \texttt{[CLS]} A \, \texttt{[SEP]} \, type(A) \\
    &\quad \texttt{[SEP]} \, v_1 \, \texttt{[SEP]} \, \ldots \, \texttt{[SEP]} \, v_m, \\
    \mathcal{S}_{\texttt{verbose}}(A) &= \texttt{[CLS]} \text{Column: } A \, \texttt{[SEP]} \, \text{Type: } type(A) \\
    &\quad \texttt{[SEP]} \, \text{Values: } v_1 \, \texttt{[SEP]} \, \ldots \, \texttt{[SEP]} \, v_m,\\
    \mathcal{S}_{\texttt{repeat}}(A) &= \texttt{[CLS]} \underbrace{A \, \texttt{[SEP]} \dots \texttt{[SEP]} \, A \, \texttt{[SEP]}}_{\text{$k$ times}} \, type(A) \\
    &\quad \texttt{[SEP]} \, v_1 \, \texttt{[SEP]} \, \ldots \, \texttt{[SEP]} \, v_m,
\end{align*}
\end{small}
where the [CLS] token is a special symbol that indicates the start of a column, and [SEP] separates column components. 

\revision{$\mathcal{S}_{\texttt{default}}$ is a variant of a widely-adopted serialization strategy~\cite{suhara2022annotating,yin2020tabert}. Instead of employing commas to separate sample values, we utilize the \texttt{[SEP]} token. This modification prevents the model from interpreting the values as an ordered text sequence, thereby treating them as unordered, discrete features. This change is crucial in schema matching as it emphasizes the structured and independent nature of the data in each column.}

\revision{$\mathcal{S}_{\texttt{verbose}}$ is an extension of $\mathcal{S}_{\texttt{default}}$ where prefixes are added to delineate each component and provide additional context for the SLM. By explicitly tagging each data segment, it helps models better contextualize the information, leading to improved interpretability and alignment accuracy in schema matching tasks.}

\revision{$\mathcal{S}_{\texttt{repeat}}$ repeats the column name multiple times to reinforce its importance, nudging the model to prioritize column names. This strategy is inspired by findings in attention-based neural network research where repetition can enhance item salience \citep{vaswani2017attention}.
It attempts to mitigate attention drift in SLMs, which can disproportionately focus on later tokens or values rather than column names, especially in zero-shot settings where the model is not optimized to learn the importance of different column components. 
For our experiments, we set \( k=5 \) to strike a balance between reinforcing column names without overwhelming the model with redundancy.}

In addition to column name and values, we incorporate data types into the column representation.  For column type inference, we classify columns into basic types: numerical, categorical, date, or binary. We classify columns with a high proportion of unique values (e.g., over 90\% distinct values) as ``key'' columns, as they often represent unique identifiers. When column value is unavailable or type detection fails, we label the column type as ``unknown''. 



In \alg, we use serialization as a \textit{hyperparameter}. We study the impact 
of different serialization strategies for schema matching in Section~\ref{sec:experiments}, where we also investigate different approaches to sampling column values.
We experimented with varying column value sample sizes (from 10 to 30) and observed only marginal differences.
Interestingly, smaller sample sizes (e.g., ten values) sometimes outperformed larger ones, likely because of reduced noise. Thus, we fixed the sample size to 10 as a default.

\begin{figure}[t]
    \centering
    \includegraphics[width=\linewidth]{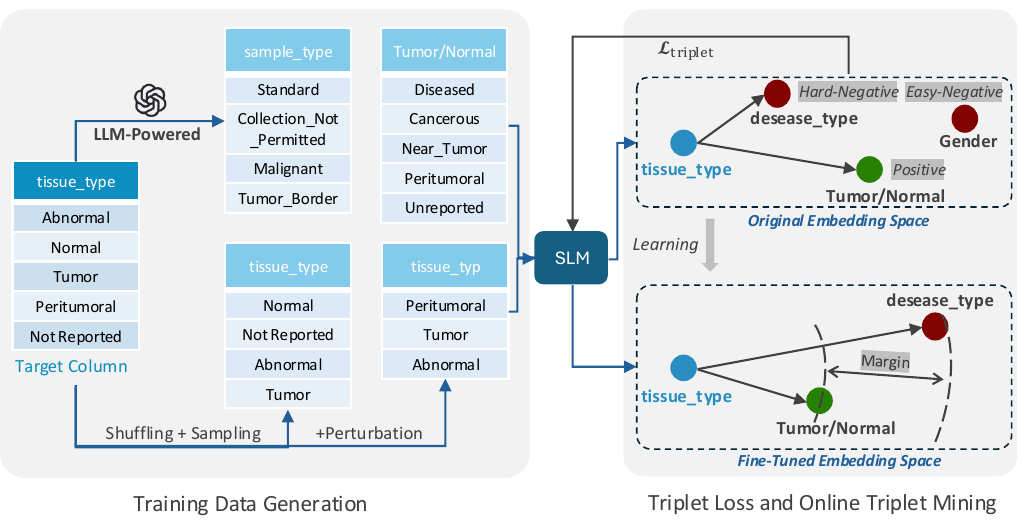}
    \vspace{-.3cm}
    \caption{LLM-Powered Fine-tuning Pipeline}
    \vspace{-.3cm}
    \label{fig:fine-tune}
\end{figure}

\subsection{LLM-Powered Fine Tuning}
\label{sec:modeltraining}

By fine-tuning, we aim to learn column representations whose embeddings cluster spatially to reflect semantic relationships. This spatial arrangement enables efficient retrieval of related columns via cosine similarity. Our training methodology thus minimizes the distance between embeddings of identical or semantically related columns while maximizing separation between distance ones.

\myparagraph{LLM-Derived Training Data}
As discussed above, an important challenge in fine-tuning lies in obtaining high-quality training data.
With the goal of capturing syntactic heterogeneity common in real datasets, we introduce \llmaugfull (\llmaug), an approach that generates variations of columns that are semantically equivalent but syntactically different. 
The derived columns serve as the training data fine-tuning process, which adopts a self-supervised contrastive learning approach (Figure~\ref{fig:fine-tune}).

Synthetic columns are derived from an input (anchor) column as follows:
1) The anchor name and a sample of its values are given to the LLM using the structured prompt depicted in Figure~\ref{fig:data-generation-prompt}, and 2) 
the LLM outputs columns that are semantically similar to the anchor. Examples of synthetic columns are shown in Table~\ref{tab:data-gen}. This method establishes a class of columns considered matches (derived from the same anchor); columns within the same class serve as positive
examples and those from different classes as negative examples.

\begin{figure}[t]
\centering
\footnotesize
\begin{tabularx}{\linewidth}{p{0.15\linewidth}X}
\toprule
\textbf{Instruction} &
\hlc[white!20]{Given the table column} \hlc[mangotango!30]{[Column Name]} \hlc[white!20]{with values} \hlc[mangotango!30]{[Sample Column Values]},
\hlc[white!20]{generate three alternative column names that adhere to typical database naming conventions such as underscores and abbreviations. 
Additionally, provide distinct, technically correct synonyms, variants, or abbreviations for the listed values. 
For columns with numerical or datetime data, generate random numbers or dates appropriate to the column's semantic meaning.}\\
\midrule
\textbf{Format} & 
\hlc[white!20]{Ensure that each set does not exceed 15 values.}
\hlc[white!20]{Format your output as follows:}
\hlc[maroon(x11)!20]{alternative\_name\_1, value1, value2, value3, ...; alternative\_name\_2, value1, value2, value3, ...; alternative\_name\_3, value1, value2, value3, ...}
\hlc[white!20]{Ensure your response excludes additional information and quotations.}\\
\bottomrule
\end{tabularx}
\caption{Training Data Generation Prompt.}
\label{fig:data-generation-prompt}
\end{figure}

\begin{table}[!ht]
\centering
\footnotesize
\caption{Examples of the provided data to LLM-based reranker and the generated data.}
\begin{tabular}{p{0.1\linewidth}p{0.37\linewidth}p{0.37\linewidth}}
\toprule
& \textbf{Original Data} & \textbf{Generated Data} \\ \midrule
\textbf{Column:} & tissue\_source\_sites & tumor\_site \\
\textbf{Values:} & \textit{Thyroid, Ovary} & \textit{Thyroidal, Ovarian} \\ \midrule
\textbf{Column:} & exon & gene\_segments \\
\textbf{Values:} & \textit{exon11, exon15} & \textit{segment11, segment15} \\ \midrule
\textbf{Column:} & masked\_somatic\_mutations & genetic\_variants \\
\textbf{Values:} & \textit{MET\_D1010N, FLT3\_ITD} & \textit{D1010N\_MET, ITD\_FLT3} \\ \midrule
\textbf{Column:} & max\_tumor\_bulk\_site & primary\_tumor\_location \\
\textbf{Values:} & \textit{Maxilla, Splenic lymph nodes} & \textit{Maxillary, Splenic\_nodes} \\ \bottomrule
\end{tabular}
\label{tab:data-gen}
\end{table}

We combine the LLM-based augmentation with other \structaugfull methods (\structaug) 
such as random sampling and shuffling of values, and minor perturbations to the column name, including random character replacements or deletions~\cite{starmie2023}. These techniques inject variability while preserving syntactic and structural similarity to the original column.
By using both augmentation strategies, the model can learn to identify matches with different characteristics.
We restrict this fine-tuning and training process to the target table columns to avoid inaccuracies. 
Applying this process to source tables with unknown true matches to target tables could mistakenly classify them as negative examples, introducing errors into the embedding space.



\myparagraph{Triplet Loss and Online Triplet Mining}
To leverage the generated synthetic columns, we implement a contrastive learning framework using \emph{triplet loss} and \emph{online triplet mining}~\citep{schroff2015facenet}. 
Unlike standard contrastive losses, triplet loss directly optimizes relative similarities, improving schema alignment precision.

Each training triplet includes an anchor column $a$, a same-class positive $p$, and a different-class negative $n$. The model learns to minimize $D(a, p)$ and maximize $D(a, n)$, where $D$ is cosine distance. 
%
Specifically, we incorporate Batch Hard triplet loss \cite{schroff2015facenet}
which is defined as follows:
\begin{equation}
\mathcal{L} = \sum_{i=1}^{P} \sum_{a=1}^{K} \big[ \max \big(0,\, m + \max_{p=1 \dots K} D(a, p) - \min_{\substack{j=1 \dots P \\ n=1 \dots K \\ j \neq i}} D(a, n) \big) \big]
\end{equation}

\noindent Here, $m$ is a margin hyperparameter that ensures a minimum distance between the positive and negative embeddings relative to the anchor, thereby enhancing the separability between classes. The loss iterates over $P$ classes and $K$ columns within each training batch, optimizing the embeddings to emphasize inter-class distinctions and intra-class similarities.

Additionally, we apply \textit{Online Triplet Mining}~\cite{schroff2015facenet,hermans2017defense}, which enhances the learning process by dynamically selecting the most challenging positive and negative examples within each training batch. This technique prioritizes triplets that maximize learning efficiency, specifically focusing on:
\begin{myitemize}
  \item \textbf{Hard Negatives}: Closest negative to the anchor that violates the margin constraint: $ n^* = \arg\min_{n} D(a, n) \text{ where } D(a, n) < D(a, p) + m$
  \item \textbf{Semi-Hard Negatives}: Negatives farther than the positive but within the margin: $D(a, p) < D(a, n) < D(a, p) + m$
  \item \textbf{Hard Positives}: Farthest positive from the anchor within its class: $p^* = \arg\max_{p} D(a, p)$
\end{myitemize}
These conditions help the model not settle for easy examples and instead learn to distinguish subtle differences, developing more robust and discriminative features to distinguish columns.
%
Together, triplet loss and mining improve embedding discriminability for schema matching.  


\myparagraph{Model Selection for Schema Matching}
To select the optimal fine-tuned model during training, we must define an effective validation metric specifically tailored for schema matching tasks.
Our selection process uses the measures described in Section~\ref{subsec:definitions}
to evaluate model performance on synthetic data. 
Since no ground-truth data is available, we rely on synthetic datasets to simulate real-world scenarios. These metrics are specifically chosen to ensure that the model identifies correct matches and ranks them in a manner that reflects their true relevance.
We compute a validation score that averages the MRR and recall at ground truth:
$
\text{Validation Score} = \big(\text{MRR} + \text{Recall@GT}\big)/2.
$
We implement early stopping when the validation accuracy remains unchanged for 5 epochs, selecting the model with the highest validation score as the best fine-tuned model.

\section{LLMs as Rerankers}
\label{sec:match_selector}
SLMs can serve as efficient retrievers, but they may fail to capture complex semantic relationships.
\hide{Moreover, the reliance on cosine similarity for scoring with SLMs may not yield accurate assessments across different source columns. This is primarily because cosine similarity measures only the angular distance between vectors, potentially overlooking deeper semantic discrepancies that vary from one source column to another.}
Moreover, models fine-tuned with LLM-derived data are inherently domain-specific.
To mitigate these limitations, \alg 
incorporates a Large Language Model (LLM) as a reranker. 
%
Unlike existing methods that adjust match rankings based on heuristic and similarity metrics~\cite{cupid2001,melnik2002similarity,coma2002, managingUncertaintyGal2006,topkGeneration2009,romeroFilteringMatches2020}, \alg leverages LLM understanding to complement initial retrieval and improve overall \smabrev performance.
%

A natural question arises: Why not just use an LLM for \smabrev? While there are approaches that employ LLMs for \smabrev~\cite{tableGPT2024, parciak2024schema}, their applicability in scenarios that involve many or large tables is limited due to high computational costs and 
challenges related to context windows (\emph{Challenge 2}, Section~\ref{sec:intro}). \revision{
We also empirically demonstrate that LLM-only approaches underperform an SLM-LLM combination for large target schemas (Figure~\ref{fig:gpt_gdc}).
} 
We posit that LLMs must be used judiciously and designed \alg accordingly. 


The core of our reranking approach is a carefully designed prompt template that converts the abstract task of column-pair similarity assessment into a more structured and interpretable process.   
We designed this prompt inspired by recent lessons learned from column type annotation~\cite{archetypeVLDB2024} and trends in prompt engineering~\cite{white2023prompt,llmWrangle2022}.
Figure \ref{fig:rerankprompt} shows the structure of the prompt.

\begin{figure}[!ht]
\centering
\footnotesize
\begin{tabularx}{\linewidth}{p{0.15\linewidth}X}
\toprule
\textbf{Scoring-Oriented Instruction} &
\hlc[white!20]{From a score of 0.00 to 1.00, judge the similarity of the candidate column in the source table to each target column in the target table. 
Each column is represented by its name and a sample of its respective values, if available.}\\
\midrule
\textbf{One-shot} \newline \textbf{Example}&
\hlc[white!20]{Example:}\newline
\hlc[white!20]{Candidate Column::}\newline
\hlc[maroon(x11)!20]{Column: \textsc{EmpID}, Sample values: \textsc{[100, 101, 102]}}\newline
\hlc[white!20]{Target Schemas:} \newline
\hlc[maroon(x11)!20]{Column: \textsc{WorkerID}, Sample values: \textsc{[100, 101, 102]}}\newline
\hlc[maroon(x11)!20]{Column: \textsc{EmpCode}, Sample values: \textsc{[00A, 00B, 00C]}} \newline
\hlc[maroon(x11)!20]{Column: \textsc{StaffName}, Sample values: \textsc{[``Alice'', ``Bob'', ``Charlie'']}}
\newline
\hlc[white!20]{Response:} \hlc[maroon(x11)!20]{\textsc{WorkerID(0.95)}; \textsc{EmpCode(0.30)}; \textsc{StaffName(0.05)}} \\
\midrule
\textbf{Format} & 
\hlc[white!20]{Provide the name of each target column followed by its similarity score in parentheses, formatted to two decimals, and separated by semicolons. Rank the column-score pairs in descending order. Exclude additional information and quotations.}\\
\midrule
\textbf{Input} & 
\hlc[white!20]{Candidate Column::}\hlc[mangotango!30]{[\textsf{Serialized Source Column}]}\newline
\hlc[white!20]{Target Schemas:}\hlc[mangotango!30]{[\textsf{Serialized Target Columns}]} \newline
\hlc[white!20]{Response:}\\
\bottomrule
\end{tabularx}
\vspace{-.3cm}
\caption{Schema Matching Prompt.}
\vspace{-.3cm}
\label{fig:rerankprompt}
\end{figure}

\myparagraph{Scoring-Oriented Prompt Design}  
The \emph{Scoring-Oriented Instruction} aligns the model with schema matching by sending one source column and the top $k$ target candidates from the SLM to the LLM. Prior works \cite{tableGPT2024, parciak2024schema} use table-wise prompts that rank columns without individual scores, limiting scalability. Column-wise ranking without scoring also hinders table-wise comparisons, such as recall@GT evaluation.
To address these limitations, we propose a novel prompt design that requires the model to assign scores from 0.00 to 1.00 to each column pair, rather than merely producing a ranked list. This facilitates direct comparisons across different pairs and allows the model to adopt a holistic view by considering all $k$ target columns simultaneously during scoring. Additionally, our serialization of the column data clarifies to the model that a text sequence comprising a column name followed by its values represents a column in a relational table. This allows more accurately scoring matches based not only on the semantic and contextual relevance but also on the structural characteristics of the data.


\myparagraph{Few-Shot In-Context Learning}
By providing a few examples of a task to an LLM, \textit{few-shot learning} can lead to significant improvement in LLM performance~\cite{brown2020language, dong2022survey}.
This method has been shown effective in various data management tasks, enabling robust performance without the need for extensive fine-tuning \cite{hegselmann2023tabllm, adnevPVLDB2020}.

In the \emph{One-shot Example} strategy, we provide a single example of the schema matching task, which outputs a ranked list with scores. This approach clarifies both the objective of the task and expected output format, and establishes a uniform scoring standard, enhancing the comparability of matches across all source and target column pairs. Utilizing only one example helps maintain a lightweight prompt structure, crucial for minimizing input and output token counts. The number of tokens directly impacts runtime and cost, and it can also influence accuracy~\cite{li-long-prompt@corr2024}.

\myparagraph{Optimization of Model Cost}
To ensure reliability, we attempt to parse the LLM's output up to three times. If parsing fails after these attempts, we revert to using the embedding scores as a fallback mechanism. During testing, output-related issues were infrequent; however, they occurred more commonly with larger tables and in scenarios lacking a candidate retriever.

Note that the design of \alg makes it possible to balance accuracy with computational cost. Based on operational constraints (or requirements), the number of candidates sent to the reranker can be adjusted. The scores for candidates not assessed by the reranker are normalized such that the maximum score aligns with the lowest score received from the reranker. Therefore, a complete ranked list can still be returned to the user. 
\section{The \GDCBenchmark}
\label{sec:benchmark}
A major challenge in evaluating schema-matching algorithms is the lack of benchmarks reflecting the diversity of real-world data.
Benchmarks are often synthetically fabricated from real data~\cite{valentine2021, magellandata}, which can introduce biases.
Moreover, as show in Section~\ref{sec:experiments} (and in~\cite{valentine2021}), publicly available benchmarks based on real data such as Magellan~\cite{magellandata} and WikiData~\cite{valentine2021} are quickly becoming ``saturated'' since many algorithms attain near-perfect performance and leave small room for algorithmic improvements. This makes it difficult to extract useful insights about the strengths of different algorithms. 

To address this problem, we built a new benchmark dataset~\cite{gdc_sm_benchmark_v1} based on the real data harmonization scenario described in Example~\ref{example:p-analysis}.
We collaborated with biomedical researchers
to design a benchmark that reflects the challenges they face when working with biomedical data. We obtained datasets from ten studies related to tumor analysis~\cite{cao2021proteogenomic, clark2019integrated, dou2020proteogenomic, gillette2020proteogenomic, mcdermott2020proteogenomic, huang2021proteogenomic, krug2020proteogenomic, satpathy2021proteogenomic, vasaikar2019proteogenomic, wang2021proteogenomic}, and, with the experts' help, manually aligned and matched these datasets to the Genomics Data Commons (GDC) standard~\cite{gdc}. 

The GDC is a program of the US National Institutes of Health responsible for handling genomic, clinical, and biospecimen data from cancer research initiatives.
Its standard dictionary describes data using a graph model that includes names and descriptions for nodes and attributes and acceptable values for some attributes. 
To be compatible with existing \smabrev solutions, we transformed the model to a relational schema that contains only column names and domain information (i.e., we disregard column descriptions).
We created a simplified table reflecting the GDC format, the ``target'' table, listing domain values for each column without repetition.

Table \ref{table:match-examples} illustrates a few samples of the data published by Dou et al. \cite{dou2020proteogenomic} alongside their corresponding GDC format. As we can observe, matching clinical data with the GDC standard poses several challenges, including terminology mismatches and data format variations. 
The benchmark includes 10 pairs of source-target tables. The number of columns in the source tables ranges from 16 to 179, and the number of rows ranges from 93 to 225. Our simplified GDC target schema comprises a single table with 736 columns. While some columns have a small number of distinct values (e.g., binary \texttt{yes/no} attributes), some contain up to 4478 distinct values.
The ground truth was manually curated by multiple annotators, who used a mix of manual and automated methods to identify possible candidate matches (e.g., GDC search tools~\cite{gdc-search-portal} and bdi-kit~\cite{bdikit}).
Given that the correctness of some matches is  challenging to determine even for bioinformatics experts (e.g., it may require reading the original papers or asking data producers), the final match decisions were made by consensus based on what users would expect from an algorithm given the limited context.
\section{Experimental Evaluation}
\label{sec:experiments}

\subsection{Experimental Setup}
\label{sec:expsetup}

\myparagraph{Datasets} 
We evaluated \alg on six datasets (Table~\ref{tab:dataset_statistics}), grouped as: 
\revision{(1) \textit{Human-Curated}-reflecting real-world matches, and (2) \textit{Fabricated}—systematically generated to capture structural variations and diverse match types.}
The \GDCBenchmark (Section \ref{sec:benchmark}) reflects
challenges in biomedical data integration.
The other five datasets are from the Valentine schema matching benchmark~\cite{valentine2021} and have been used to evaluate recent schema matching solutions~\cite{tu2023unicorn,inSituSchMatcICDE2024}. 

\begin{table}[t]
\centering
\caption{Statistics of the datasets used for experiments.}
\label{tab:dataset_statistics}
\begin{small}
\begin{tabular}{lcccc}
\toprule
\textbf{Dataset} & \makecell{\textbf{\#Table}\\\textbf{Pairs}} & \textbf{\#Cols} & \textbf{\#Rows} & \textbf{\revision{Match Type}} \\ 
\midrule
\hlc[bleudefrance!30]{\textbf{\GDC}}                 & \hlc[bleudefrance!30]{10}                  & \hlc[bleudefrance!30]{16--736}   & \hlc[bleudefrance!30]{93--4.5k}      & \revision{\hlc[bleudefrance!30]{Human-Curated}}      \\
Magellan            & 7                  & 3--7         & 0.9k--131k    & \revision{Human-Curated}       \\ 
WikiData            & 4                  & 13--20       & 5.4k--10.8k   & \revision{Human-Curated}       \\ 
Open Data           & 180                & 26--51       & 11.6k--23k    & \revision{Fabricated}       \\ 
ChEMBL              & 180                & 12--23       & 7.5k--15k     & \revision{Fabricated}       \\ 
TPC-DI              & 180                & 11--22       & 7.5k--15k     & \revision{Fabricated}  \\ 
\bottomrule
\end{tabular}
\end{small}
\end{table}

\myparagraph{Baselines} 
We compare \alg against several approaches, ranging from traditional string-similarity-based methods to sophisticated model-driven techniques. 
For traditional approaches, we use:
\begin{myitemize}
    \item \texttt{COMA}: Combines multiple strategies to compute and aggregate the similarity of table metadata~\cite{coma2002};
    \item \texttt{COMA++}: An extension of \texttt{COMA} that leverages column values and their distributions~\cite{comaplusplus2005};
    \item \texttt{Distribution}: Detects column correspondences based on data value distribution ~\cite{attributediscovery2011};
    \item \texttt{SimFlooding}: Uses several graph-based techniques to identify correspondences between metadata~\cite{melnik2002similarity}.
\end{myitemize}

\noindent We evaluated the \texttt{Jaccard-Levenshtein} baseline proposed in~\cite{valentine2021} and the  \texttt{Cupid} algorithm~\cite{cupid2001}. However, both methods consistently performed significantly worse than other approaches across all experiments, so we excluded them from the plots for clarity. 
We used the implementations available in the Valentine repository for all traditional approaches~\cite{valentine2021}.

We also compared \alg against recent approaches that use LMs:
\texttt{ISResMat} which leverages contrastive learning and embeddings from SLMs~\cite{inSituSchMatcICDE2024}, and
\texttt{Unicorn}, a supervised and general approach using language models for data encoding and trained for data integration tasks~\cite{tu2023unicorn}.  
\hide{\begin{myitemize}
\item \texttt{ISResMat}: Leverages contrastive learning and embeddings from SLMs~\cite{inSituSchMatcICDE2024}.
\item \texttt{Unicorn}: A supervised and general approach using language models for data encoding and trained for data integration tasks~\cite{tu2023unicorn}.  
\end{myitemize}}
\noindent We used the implementation from the authors for \texttt{ISResMat}.
For \texttt{Unicorn}, we used the pre-trained model and implementation provided by the authors. The match scores were derived from the model's predicted match probabilities. It is important to note that \texttt{Unicorn} requires external training data for optimal performance. The zero-shot version performed significantly worse than other baselines and was excluded from our evaluation.
We report Unicorn numbers for the algorithm in Valentine datasets for completeness, but note that its model was trained on the fabricated datasets from Valentine~\cite{valentine2021}, so data leakage can influence the reported results.

\myparagraph{Implementation Details}
The \coma algorithm is implemented in Java. \alg and all other methods are written in Python.
We use MPNet as our underlying small language model. It is pre-trained on masked and permuted language tasks, enabling precise contextual understanding of unordered text in column headers and entries~\cite{song2020mpnet,deepJoin2023,sheetrit2024rematch}. For the LLM reranker, we use the \texttt{GPT-4o-mini} model from the \texttt{OpenAI} API due to its robust performance and cost-effectiveness. However, these choices are flexible; alternative models can be easily integrated as replacements based on cost or performance requirements. It is important to note that our study does not focus on determining the optimal model for this task.

We assess four variations of \alg that represent different combinations of retrievers and rerankers: \algzsbp (zero-shot SLM, bipartite reranker), \algftbp (fine-tuned SLM, bipartite reranker), \algzsgpt (zero-shot SLM, LLM reranker),
\algftgpt (fine-tuned SLM, LLM retriever).

We set margin $m=0.5$ for the fine-tuning configuration, and run the model for 30 epochs on the \GDCBenchmark and 10 epochs on the Valentine datasets.


\myparagraph{Execution platform}
The experiments were run on a server running Red Hat Enterprise Linux version 9.2. 
We used a single node consisting of 8 cores and 64 GB of memory for CPU-based tasks. 
The GPU-intensive experiments used an NVIDIA A100 GPU.
For \alg variations, Unicorn, and IsResMat, we used a GPU to run all experiments, except the scalability experiment described in Section \ref{sec:main_scalability}, conducted on a CPU for consistency with other methods.

\begin{figure}[t]
  \centering
  \includegraphics[width=0.8\columnwidth]{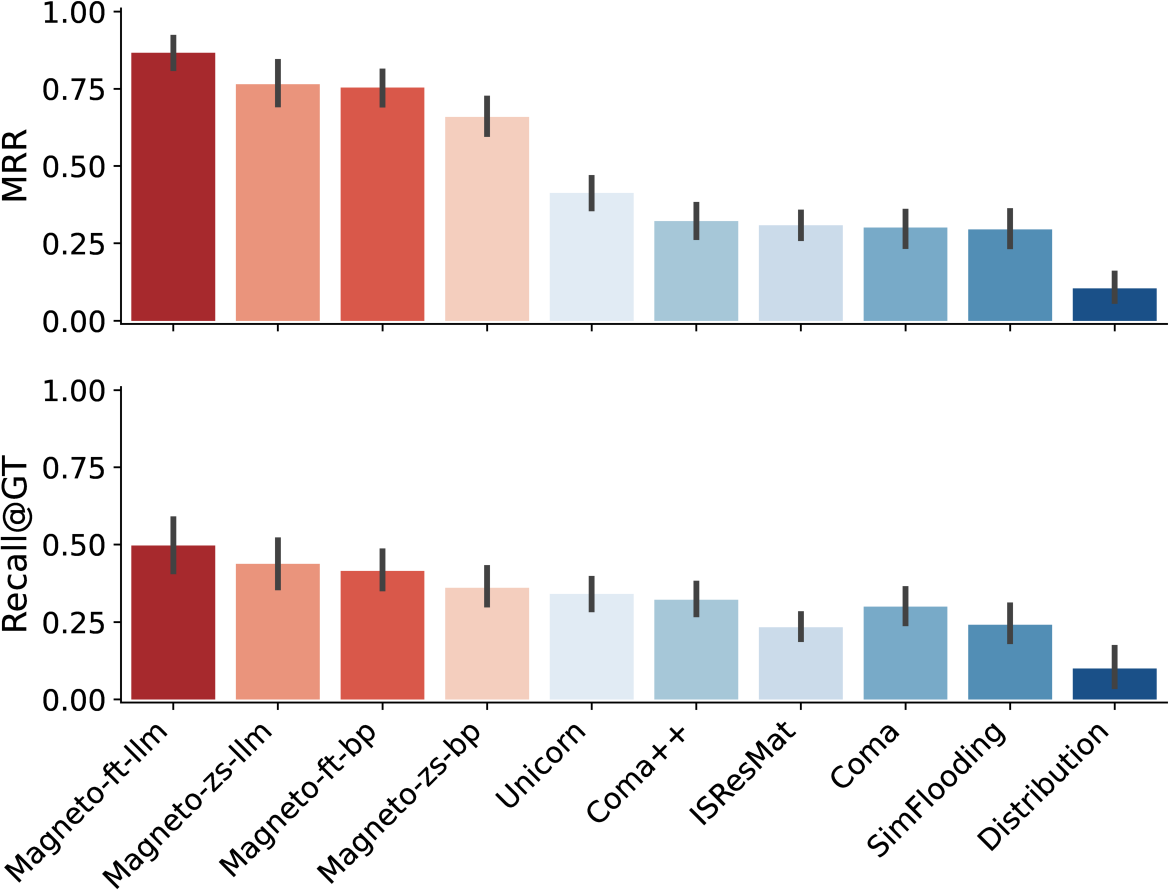}
\vspace{-.2cm}
\caption{\alg attains higher accuracy on the challenging \GDCBenchmark than traditional, model-based, and supervised approaches to SM.}
  \vspace{-.4cm}
  \label{fig:compbaselines_gdc}
\end{figure}

\subsection{End-to-End Accuracy}
\label{secmain_accuracy}

\myparagraph{Performance on \GDC}
%
Figure~\ref{fig:compbaselines_gdc} shows the accuracy (measured by MRR and Recall@GT) for all versions of \alg and the baseline methods for the \GDCBenchmark.
The \alg variations outperform all other methods for both measures. \algftgpt has the highest overall accuracy, confirming that the combination of a  fine-tuned SLM and an LLM is effective for this complex matching task.
The fact that the zero-shot \alg variations also outperform recent state-of-the-art approaches demonstrates the effectiveness of the serialization and sampling techniques we designed for schema matching. We ablate the different components of \alg in Section~\ref{sec:in-depth}, where we discuss this in more detail.

\begin{figure}[t]
  \centering
  \includegraphics[width=0.9\columnwidth]{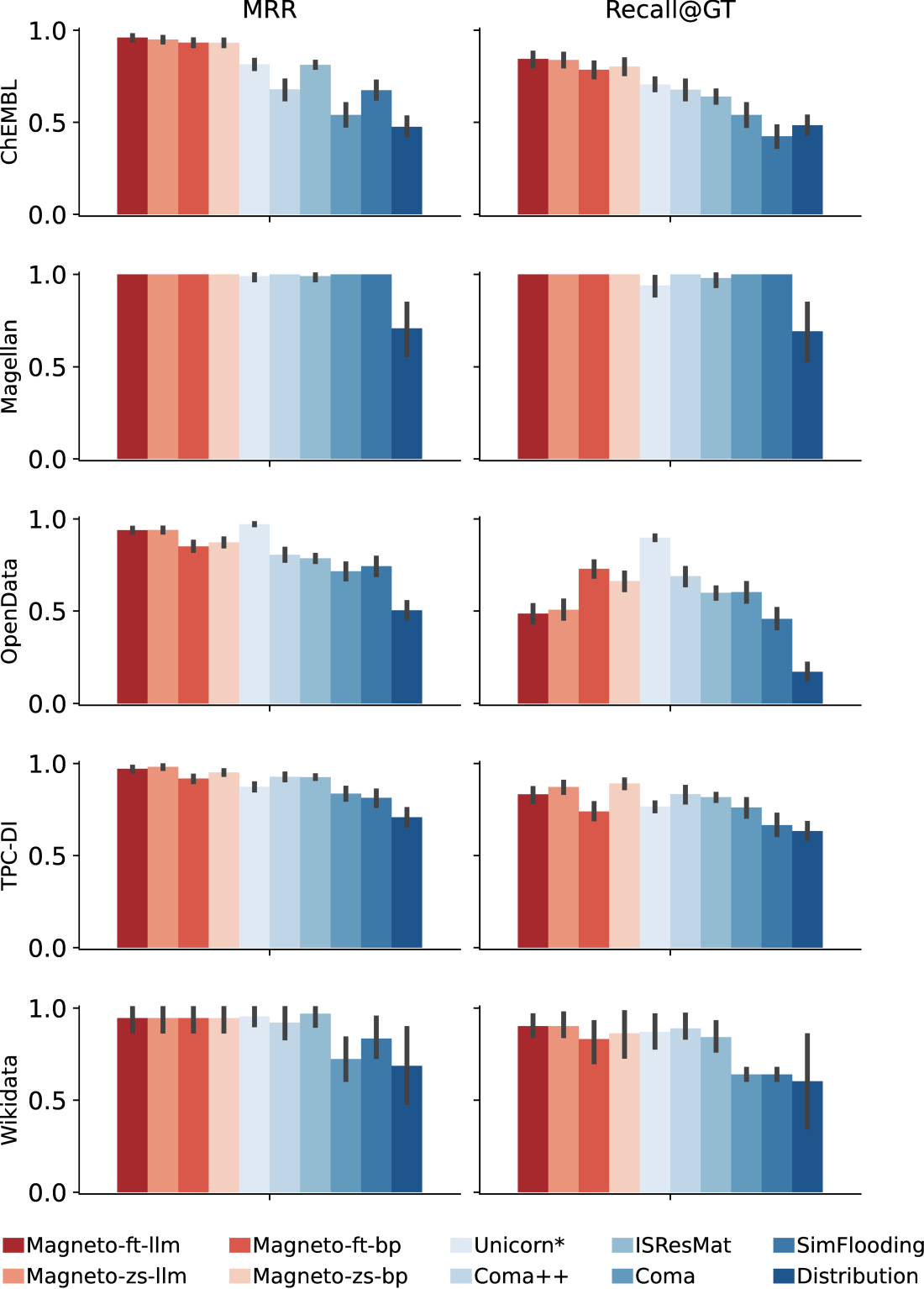}
\vspace{-.2cm}
\caption{Accuracy on  Valentine datasets. 
\alg variants attain high accuracy (even without fine-tuning) compared to traditional, model-based, and supervised approaches.}
\vspace{-.4cm}

  \label{fig:compbaselines_valentine}
\end{figure}

\myparagraph{Performance on Valentine} 
The accuracy results for the Valentine benchmark are shown in Figure \ref{fig:compbaselines_valentine}. Note that the different methods attain much higher accuracies for these datasets than for \GDC. 
The Valentine datasets are derived either from open data sources or synthetic repositories, and they are also meticulously curated. 
In contrast, the \GDCBenchmark includes real scientific datasets and features a diverse set of column names and values, which,  as these results confirm, is particularly challenging for \smabrev solutions.

For all Valentine datasets, \alg performs on par with or better than the baselines. 
The only exception is \texttt{Unicorn}. 
However, as previously noted, this can likely be attributed to data leakage: \texttt{Unicorn} is trained using Valentine datasets.
This advantage is particularly apparent in datasets like OpenData: \texttt{Unicorn} achieves significantly higher recall than competing methods. Conversely, for the Magellan dataset, which was not used to train \texttt{Unicorn}, \texttt{Unicorn} has lower accuracy than the other methods, including methods that use only basic metadata, which achieve perfect scores.

The high accuracy of the zero-shot configurations of \texttt{\alg} leaves little room for improvement. Specifically, fine-tuning faces challenges within the Valentine datasets, which predominantly contain lexical rather than semantic matches. For instance, the OpenData dataset includes matches like \texttt{Gender} to \texttt{Ge} and \texttt{Employer} to \texttt{Em}, which are uncommon in practical applications. Despite these obstacles, 
our LLM-based fine-tuning and re-ranking generally enhance performance.
Additionally, we demonstrate in Section~\ref{sec:exp-reranker} that combining simple synthetic data generation techniques with LLM-derived data leads to better performance in this context.

\subsection{Scalability Assessment}
\label{sec:main_scalability}
We compare the runtime of \texttt{\alg} to those of the top three baselines in accuracy: \texttt{Unicorn}, \texttt{Coma++}, and \texttt{ISResMat}. Since \texttt{Coma++} does not support GPU execution, we ran this experiment in CPU mode to ensure a fair comparison.
We report only the results of \algzsbp and \algzsgpt since their runtime is similar to \algftbp and \algftgpt, respectively.

\begin{figure}[t]
  \centering
  \includegraphics[width=.9\columnwidth]{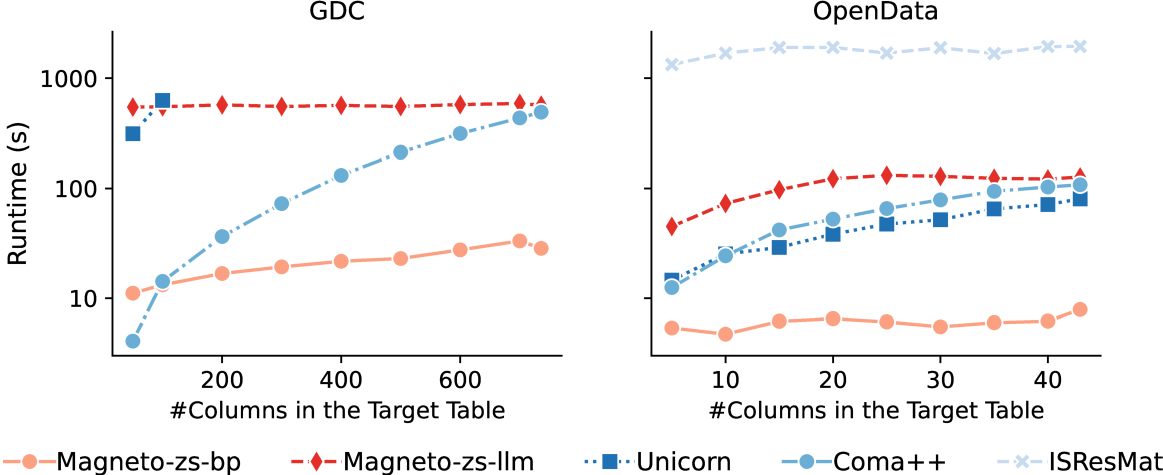}
\caption{
Runtime analysis (log-scale y-axis). \alg scales well with large datasets: \algzsbp is often much faster than baselines, while \algzsgpt maintains consistent runtimes, finishing large tasks within minutes. ISResMat and Unicorn fail to complete \GDC after 100 columns.
}
  \label{fig:runtime}
\end{figure}

For this experiment, we focus on datasets featuring tables with a large number of columns or rows, and we select one source-target pair each from the \GDC and OpenData. For the \GDC dataset, the source table comprises 179 rows and 153 columns, while the target table contains 4.5k rows and 733 columns. For OpenData, both the source and the target table contain 23k rows and 43 columns. 
We maintain the input table static and incrementally increase the number of target columns using a random selection. Each execution is repeated 10 times per column number to accommodate randomness.
We used a time limit of 2 hours per execution and canceled the executions of any method that exceeded this limit.
The results are shown in Figure \ref{fig:runtime}.

\algzsbp and \algzsgpt remain stable with increasing table size: the runtime for \algzsbp grows slightly and \algzsgpt maintains a stable runtime despite its complexity.
For \GDC, \algzsbp shows runtimes ranging from 11--33 seconds.
Such a low increase in runtime reflects its efficient design, bounded primarily by the embedding computations.
As expected, \algzsgpt incurs higher runtimes due to the LLM requests (545--589 seconds), but it shows stability as the number of columns increases 
since the amount of data sent does not change.
The runtimes of \alg variations are even lower for OpenData: the dataset has fewer columns but many more rows, which \alg compensates through sampling.

\texttt{Coma++} shows low runtimes for a small number of columns, but its performance decreases as the number of columns grows. 
\texttt{IsResMat} and \texttt{Unicorn} exhibit significant scalability challenges, as their runtime grows substantially with the number of columns.
\texttt{IsResMat} was not able to complete the execution for \GDC, not even for the lowest number of columns, and its runtime for OpenData was orders of magnitude higher than the other methods. \texttt{Unicorn} could only process the initial 100 columns for \GDC.

\subsection{Ablation of \alg Components}
\label{sec:in-depth}

\begin{figure}[t]
    \centering
\includegraphics[width=0.9\columnwidth]{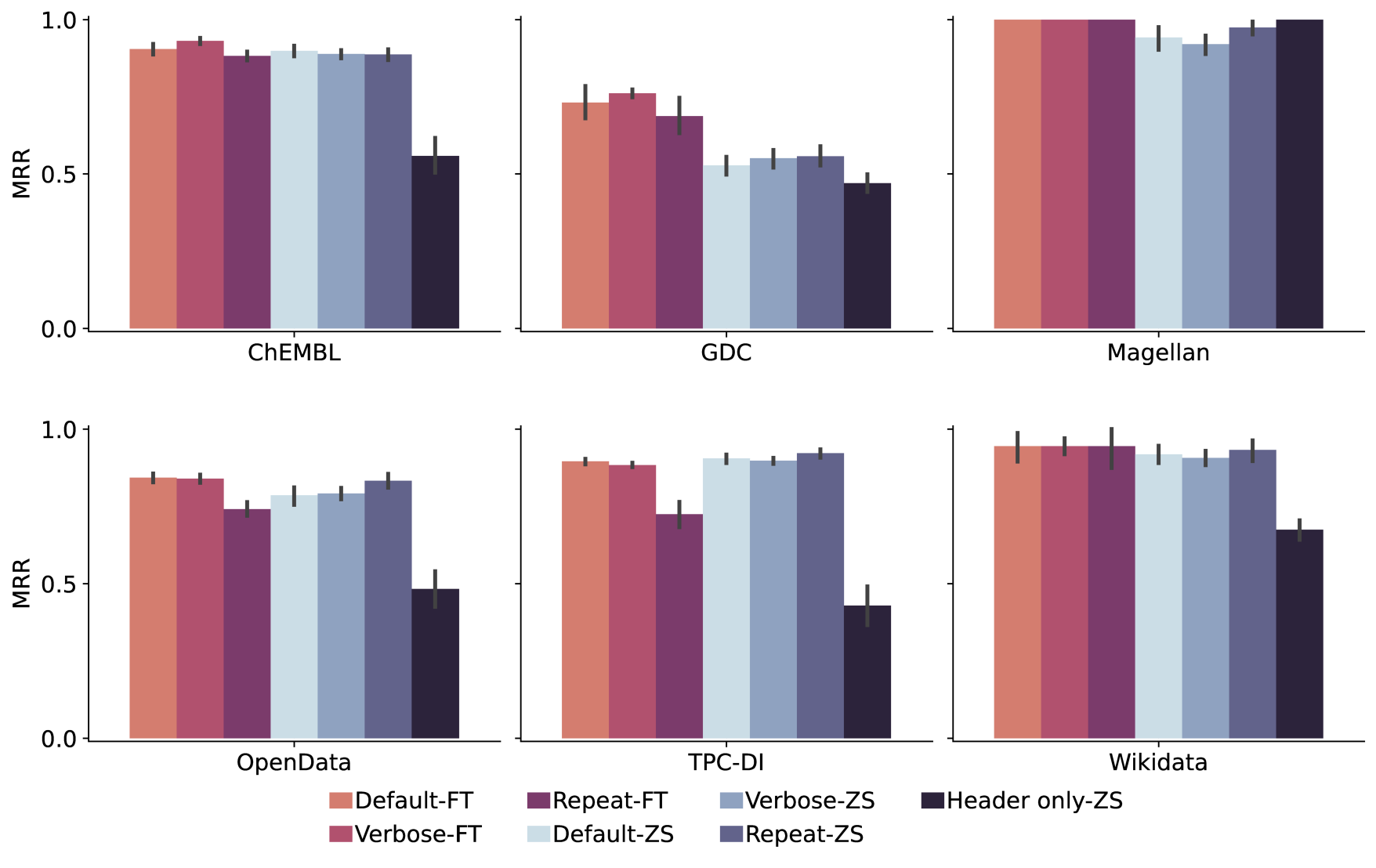}
\vspace{-.2cm}
\caption{
Ablation of column serialization strategies.
}
\vspace{-.4cm}
\label{fig:serialization_ablations}
\end{figure}

\myparagraph{Column Serialization} 
Figure~\ref{fig:serialization_ablations} shows the impact of the serialization methods (Section~\ref{sec:serialization_sampling}) on schema matching performance across all datasets based on MRR. We used the serialization strategies on fine-tuned (\texttt{FT}) and zero-shot (\texttt{ZS}) settings: Default, using $\mathcal{S}_{\texttt{default}}$, which combines column names with a small sample of values; Verbose, which uses $\mathcal{S}_{\texttt{verbose}}$ and enriches $\mathcal{S}_{\texttt{default}}$ by incorporating additional instructions; Repeat, which emphasizes header names with $\mathcal{S}_{\texttt{repeat}}$. We also include \texttt{Header only-ZS}, which uses only column headers without values, serving as a baseline. 

\revision{In zero-shot settings, $\mathcal{S}_{\texttt{repeat}}$ consistently outperforms other methods, demonstrating the effectiveness of emphasizing the column header and raising its cumulative attention in the zero-shot regime.   
While $\mathcal{S}_{\texttt{verbose}}$ under-performs in zero-shot settings, it excels after fine-tuning because the prefixes act as learnable anchors that guide the model to separate and re-weight heterogeneous components (name, type, values) during representation learning, highlighting the benefits of integrating semantic details when domain-specific training is applied.}
The \texttt{Header only-ZS} approach shows the lowest performance in all datasets, confirming that column headers alone are insufficient in complex scenarios. 

The results further support our assumption that fine-tuning can greatly improve the performance gains on the \GDC dataset. This improvement is likely related to the requirement for external knowledge, which we capture with the LLM-derived training data.

\myparagraph{Value Sampling} 
We compare the effectiveness our proposed \textit{Priority Sampling} (Section~\ref{sec:serialization_sampling})
against the following sampling methods: \texttt{Coordinated}, a variation of \texttt{Priority} that excludes frequency weights; \texttt{Weighted} and \texttt{Frequency}, variations of \texttt{Priority} without coordination but with value frequencies --  \texttt{Frequency} uses the most frequent values and \texttt{Weighted} uses weighted sampling based on these frequencies; \texttt{Random} uses basic random sampling. 

We used \GDC and the three synthetic datasets from Valentine; we exclude Wikidata and Magellan since as all methods already perform well on them.
As the results in Table~\ref{tab:sampling} show,
\algzsbp with $\mathcal{S}_{\texttt{repeat}}$, the best zero-shot setting configuration, \texttt{Priority}, generally outperforms other methods in both Recall@GT and MRR, and sometimes is a close second.
%
Priority Sampling prioritizes frequently occurring values and enhances the likelihood of sampling similar values across columns, a beneficial feature for schema matching. The top three techniques—including \texttt{Priority} and its ablations—perform comparably well, also representing viable choices.

\myparagraph{Synthetic Data Generation} 
We evaluate our \llmaug
for SLM fine-tuning against two other methods: \structaug, which incorporates row shuffling, sampling, and column name perturbation, and \texttt{mixed-aug}, which combines these methods. 
We used the optimal settings from serialization studies, $\mathcal{S}_{\texttt{repeat}}$, on \algftbp and \algftgpt. 
We assess MRR for both methods and Recall@GT for \algftbp, as Recall@GT for \algftgpt is largely influenced by the reranker instead of the fine-tuned candidate retriever. 
This analysis focuses on the four datasets that have the potential for improvement.
Table~\ref{tab:exp_data_gen} shows that \llmaug and \texttt{mixed-aug} outperform simple perturbations on the \GDC dataset. However, on Valentine datasets, LLM-generated data has limited impact due to sparse, uncommon matches. Still, the strong \GDC results highlight the effectiveness of our fine-tuning for real data.


\begin{table}[t]
\centering
\scriptsize
\caption{
Ablation of sampling techniques using \algzsbp with $\mathcal{S}_{\texttt{repeat}}$. 
\texttt{Priority} generally outperforms other techniques in Recall@GT and MRR metrics.
The best-performing techniques are highlighted in dark blue \hlc[bleudefrance!50]{\;\;\;} , with the second best in light blue \hlc[bleudefrance!20]{\;\;\;}.} 
%
%
\label{tab:sampling}
\resizebox{\columnwidth}{!}{%
\begin{tabular}{@{}lcccc@{}}
\toprule
\makecell{\textbf{Sampling}\\\textbf{Method}}& \textbf{\GDC} & \textbf{ChEMBL} & \textbf{OpenData} & \textbf{TPC-DI} \\ \midrule
\multicolumn{5}{c}{\textbf{Recall@GT}} \\ \midrule
Priority       & \hlc[bleudefrance!50]{\textbf{0.344±0.081}} & \hlc[bleudefrance!50]{\textbf{0.620±0.264}} & \hlc[bleudefrance!50]{\textbf{0.543±0.294}} & \hlc[bleudefrance!50]{\textbf{0.726±0.174}} \\
Coordinated    & 0.336±0.069 & 0.601±0.260 & 0.506±0.292 & 0.675±0.224 \\
Weighted       & \hlc[bleudefrance!20]{0.342±0.070} & \hlc[bleudefrance!20]{0.603±0.266} & 0.497±0.291 & 0.643±0.210 \\
Frequency      & 0.332±0.062 & 0.525±0.296 & \hlc[bleudefrance!20]{0.526±0.300} & \hlc[bleudefrance!20]{0.692±0.196} \\
Random         & 0.334±0.075 & 0.572±0.268 & 0.489±0.282 & 0.665±0.197 \\ \midrule
\multicolumn{5}{c}{\textbf{MRR}} \\ \midrule
Priority       & \hlc[bleudefrance!20]{0.591±0.094} & \hlc[bleudefrance!20]{0.900±0.103} & \hlc[bleudefrance!50]{\textbf{0.847±0.200}} & \hlc[bleudefrance!50]{\textbf{0.948±0.082}} \\
Coordinated    & 0.586±0.106 & \hlc[bleudefrance!50]{\textbf{0.902±0.105}} & \hlc[bleudefrance!20]{0.837±0.196} & \hlc[bleudefrance!20]{0.937±0.087} \\
Weighted       & \hlc[bleudefrance!50]{\textbf{0.599±0.099}} & 0.888±0.104 & 0.823±0.212 & 0.930±0.091 \\
Frequency      & 0.577±0.104 & 0.851±0.163 & 0.833±0.202 & 0.886±0.130 \\
Random         & 0.579±0.096 & 0.885±0.106 & 0.830±0.195 & 0.918±0.101 \\ 
\bottomrule
\end{tabular}%
}
\end{table}

\begin{table}[t]
\centering
\scriptsize
\caption{Ablation of data generation techniques. LLM-powered data generation is effective, and better performance can be achieved by combining multiple techniques.}
\label{tab:exp_data_gen}
\resizebox{\columnwidth}{!}{%
\begin{tabular}{@{}lcccc@{}}
\toprule
\makecell{\textbf{Data}\\\textbf{Generation}}& \textbf{\GDC} & \textbf{ChEMBL} & \textbf{OpenData} & \textbf{TPC-DI} \\ \midrule
\multicolumn{5}{c}{\textbf{Recall@GT for \algftbp}} \\ \midrule
\llmaug      & 0.414±0.106 & \hlc[bleudefrance!50]{\textbf{0.785±0.263}} & \hlc[bleudefrance!20]{0.729±0.271} & 0.740±0.298 \\
\texttt{mixed-aug}  & \hlc[bleudefrance!50]{\textbf{0.438±0.085}} & \hlc[bleudefrance!20]{0.774±0.273} & \hlc[bleudefrance!50]{\textbf{0.743±0.255}} & \hlc[bleudefrance!20]{0.763±0.255} \\
\structaug           & \hlc[bleudefrance!20]{0.418±0.099} & 0.764±0.282 & 0.711±0.271 & \hlc[bleudefrance!50]{\textbf{0.773±0.261}} \\ \midrule
\multicolumn{5}{c}{\textbf{MRR for \algftbp}} \\ \midrule
\llmaug        & \hlc[bleudefrance!20]{0.754±0.093} & \hlc[bleudefrance!50]{\textbf{0.932±0.103}} & \hlc[bleudefrance!50]{\textbf{0.851±0.152}} & \hlc[bleudefrance!50]{\textbf{0.917±0.100}} \\
\texttt{mixed-aug}  & \hlc[bleudefrance!50]{\textbf{0.761±0.071}} & \hlc[bleudefrance!20]{0.931±0.101} & \hlc[bleudefrance!20]{0.841±0.162} & 0.885±0.125 \\
\structaug           & 0.731±0.099 & 0.927±0.103 & 0.817±0.181 & \hlc[bleudefrance!20]{0.905±0.117} \\ \midrule
\multicolumn{5}{c}{\textbf{MRR for \algftgpt}} \\ \midrule
\llmaug        & \hlc[bleudefrance!50]{\textbf{0.866±0.083}} & \hlc[bleudefrance!50]{\textbf{0.960±0.089}} & \hlc[bleudefrance!50]{\textbf{0.939±0.080}} & \hlc[bleudefrance!50]{\textbf{0.971±0.079}} \\
\texttt{mixed-aug}  & \hlc[bleudefrance!20]{0.830±0.077} & 0.949±0.109 & \hlc[bleudefrance!20]{0.927±0.096} & 0.965±0.096 \\
\structaug           & 0.798±0.116 & \hlc[bleudefrance!20]{0.955±0.097} & 0.917±0.105 & \hlc[bleudefrance!20]{0.969±0.081} \\ 
\bottomrule
\end{tabular}%
}
\end{table}

\smallskip\noindent\textbf{\revision{Model Ablations.}}
\revision{
We extended our ablation studies to include RoBERTa~\citep{liu2019roberta} and E5 \citep{wang2022text} (SLMs), and LLaMA3.3-70B \citep{dubey2024llama} (LLM).  Table~\ref{tab:model-ablation} shows the results. 
Regarding SLMs the default MPNet outperforms other SLMs in most variations of \alg. In general, the serialization and score-based prompt proposed are robust. While $\mathcal{S}_{\texttt{repeat}}$ dominates most zero-shot settings, $\mathcal{S}_{\texttt{verbose}}$ and  $\mathcal{S}_{\texttt{default}}$ yield even higher gains after fine-tuning, which is consistent with our findings in Figure~\ref{fig:serialization_ablations}. 
The tested models benefit from the same prompting framework, which confirms that \alg performs consistently using different models with a robust cost-accuracy tradeoff.
For LLMs, LLaMA3.3-70B surpasses GPT-4o-mini (MRR=0.837 vs. 0.815), confirming that larger LLMs enhance reranking. Combining LLaMA3.3-70B with fine-tuned MPNet achieves the best overall performance (MRR=0.860), reinforcing the value of having a pipeline that performs retrieval and reranking.
}

\begin{table*}[t]
\centering
\caption{
\revision{Model ablation comparing three SLMs and two LLMs with MRR$/$Recall@GT metrics. Bold indicates best serialization per model; \hlc[bleudefrance!20]{\;\;\;} shows best zero-shot model and \hlc[bleudefrance!50]{\;\;\;} best fine-tuned model.}
}
\label{tab:model-ablation}
\resizebox{\textwidth}{!}{%
\begin{tabular}{lcccccc}
\toprule			
\textbf{Model} & \textbf{zs-bp} & \textbf{ft-bp} & \textbf{zs-gpt4o-mini} & \textbf{ft-gpt4o-mini} & \textbf{zs-llama3.3-70b} & \textbf{ft-llama3.3-70b} \\
\midrule
\textbf{MPNet ($\mathcal{S}_{\texttt{default}}$)} & 
  0.656$\pm$0.082 / 0.357$\pm$0.093 & 
  0.740$\pm$0.089 / 0.400$\pm$0.094 &
  0.775$\pm$0.102 / 0.412$\pm$0.115 &
  \textbf{\hlc[bleudefrance!50]{0.846$\pm$0.104 / 0.537$\pm$0.162}} &
  0.772$\pm$0.101 / 0.339$\pm$0.155 &
  \textbf{\hlc[bleudefrance!50]{0.860$\pm$0.088} / 0.423$\pm$0.170} \\
\textbf{MPNet ($\mathcal{S}_{\texttt{verbose}}$)} & 
  0.627$\pm$0.119 / 0.341$\pm$0.114 & 
  \textbf{\hlc[bleudefrance!50]{0.761$\pm$0.071 / 0.438$\pm$0.085}} &
  0.758$\pm$0.118 / \textbf{\hlc[bleudefrance!20]{0.469$\pm$0.131}} &
  0.830$\pm$0.766 / 0.479$\pm$0.154 &
  0.776$\pm$0.098 / 0.323$\pm$0.153 &
  0.838$\pm$0.121 / \textbf{\hlc[bleudefrance!50]{0.446$\pm$0.185}} \\
\textbf{MPNet ($\mathcal{S}_{\texttt{repeat}}$)} & 
  \textbf{\hlc[bleudefrance!20]{0.731$\pm$0.086 / 0.375$\pm$0.108}} & 
  0.701$\pm$0.110 / 0.172$\pm$0.103 &
  \textbf{\hlc[bleudefrance!20]{0.808$\pm$0.095}} / 0.430$\pm$0.131 &
  0.820$\pm$0.117 / 0.524$\pm$0.104 &
  \textbf{\hlc[bleudefrance!20]{0.828$\pm$0.096} / 0.358$\pm$0.117} &
  0.819$\pm$0.102 / 0.419$\pm$0.138 \\
\textbf{RoBERTa ($\mathcal{S}_{\texttt{default}}$)} & 
  0.631$\pm$0.105 / 0.336$\pm$0.081 & 
 \textbf{ 0.703$\pm$0.120 / 0.378$\pm$0.107} &
  0.734$\pm$0.119 / 0.381$\pm$0.088 &
  0.784$\pm$0.131 / 0.490$\pm$0.149 &
  0.755$\pm$0.098 / 0.292$\pm$0.119 &
  0.815$\pm$0.121 / \textbf{0.432$\pm$0.164} \\
\textbf{RoBERTa ($\mathcal{S}_{\texttt{verbose}}$)} & 
  0.621$\pm$0.063 / 0.350$\pm$0.083 & 
  0.692$\pm$0.114 / 0.367$\pm$0.112 &
  0.743$\pm$0.097 / 0.400$\pm$0.099 &
  \textbf{0.821$\pm$0.092 / 0.511$\pm$0.160} &
  0.751$\pm$0.075 / 0.324$\pm$0.140 &
  \textbf{0.854$\pm$0.101} / 0.424$\pm$0.147 \\
\textbf{RoBERTa ($\mathcal{S}_{\texttt{repeat}}$)} & 
  \textbf{0.710$\pm$0.103 / 0.373$\pm$0.097} & 
  0.681$\pm$0.096 / 0.368$\pm$0.111 &
  \textbf{0.794$\pm$0.097 / 0.434$\pm$0.173} &
  0.774$\pm$0.101 / 0.504$\pm$0.161 &
  \textbf{0.814$\pm$0.088 / 0.398$\pm$0.161} &
  0.810$\pm$0.090 / 0.428$\pm$0.165 \\
\textbf{E5 ($\mathcal{S}_{\texttt{default}}$)} & 
  0.623$\pm$0.104 / 0.329$\pm$0.105 & 
  \textbf{0.729$\pm$0.081 / 0.406$\pm$0.094} &
  0.738$\pm$0.144 / 0.356$\pm$0.144 &
  \textbf{0.832$\pm$0.080} / 0.480$\pm$0.117 &
  0.756$\pm$0.120 / \textbf{\hlc[bleudefrance!20]{0.405$\pm$0.090}} &
  \textbf{0.857$\pm$0.076} / 0.388$\pm$0.112 \\
\textbf{E5 ($\mathcal{S}_{\texttt{verbose}}$)} & 
  0.602$\pm$0.094 / 0.322$\pm$0.102 & 
  0.692$\pm$0.114 / 0.367$\pm$0.112 &
  0.745$\pm$0.131 / \textbf{0.382$\pm$0.135} &
  0.821$\pm$0.092 / \textbf{0.511$\pm$0.160} &
  0.746$\pm$0.100 / 0.327$\pm$0.100 &
  0.838$\pm$0.121 / \textbf{\hlc[bleudefrance!50]{0.446$\pm$0.185}} \\
\textbf{E5 ($\mathcal{S}_{\texttt{repeat}}$)} & 
  \textbf{0.715$\pm$0.120 / 0.372$\pm$0.100} & 
  0.701$\pm$0.110 / 0.372$\pm$0.103 &
  \textbf{0.797$\pm$0.121} / 0.379$\pm$0.112 & 
  0.813$\pm$0.113 / 0.502$\pm$0.154 &
  \textbf{0.812$\pm$0.118} / 0.325$\pm$0.144 &
  0.810$\pm$0.090 / 0.428$\pm$0.165 \\
\bottomrule
\end{tabular}
} 
\end{table*}

\begin{figure}[t]
    \centering
\includegraphics[width=0.8\columnwidth]{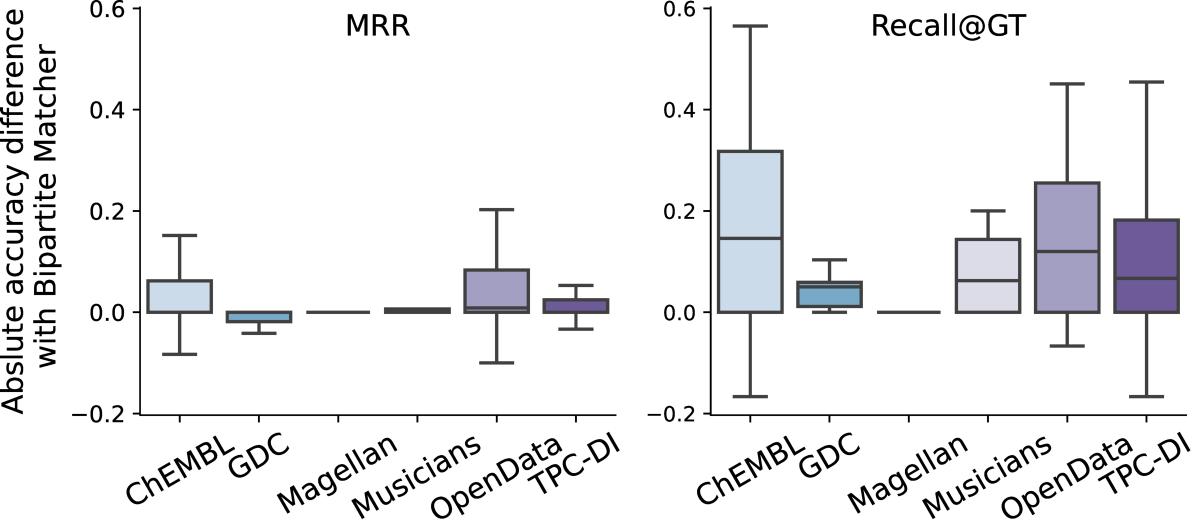}
\caption{Ablation of rerankers. Absolute accuracy improvement using \algzsbp over \alg-zs.}
    \label{fig:bipartite}
\end{figure}

\smallskip\noindent\textbf{Match Reranker.}
\label{sec:exp-reranker}
We first examine the impact of using the bipartite strategy (Section~\ref{subsec:overview}). Figure~\ref{fig:bipartite} shows the increase/decrease in accuracy from \alg-zs, which directly returns the retriever matches, to \algzsbp.
%
Note that the bipartite approach can significantly improve Recall@GT. All datasets benefit from the technique at some point, except Magellan, whose scores are already very high. The absolute increase can reach nearly 0.6. 
The improvements for MRR are lower but still sizable for ChEMBL and OpenData. 

\begin{figure}[t]
\centering
\includegraphics[width=.8\columnwidth]{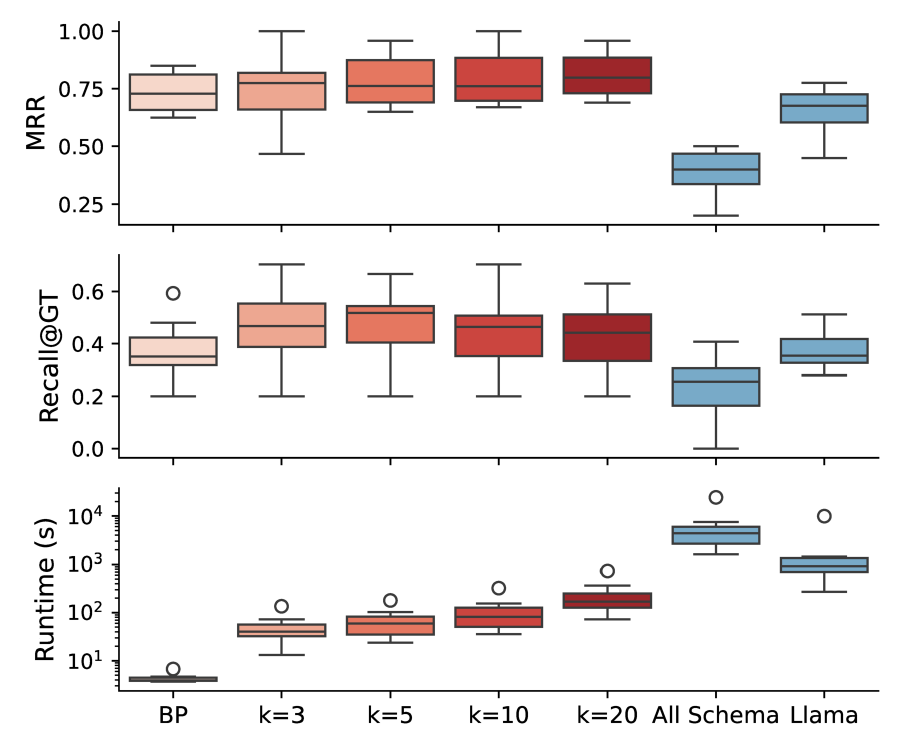}
\caption{\revision{Ablation of candidate counts ($k=3$ to $20$) sent to GPT-4o-mini in \algzsgpt. 
We also compare baselines that use \algzsbp (BP) for reranking, and use only an LLM -- and LLaMA3.3-70B  (Llama) and GPT-4o-mini (All Schema).
Runtime shown on a \textit{logarithmic} scale.}}
\label{fig:gpt_gdc}
\end{figure}

\revision{We also evaluate: (1) improvements of the LLM-based reranker over the bipartite method, (2) performance variations with more candidates sent to the LLM, and (3) outcomes when bypassing the SLM retrieval stage. We focus on the \GDC dataset, noted for its high column count, using zero-shot settings for the retriever. Figure \ref{fig:gpt_gdc} presents accuracy and runtime for different numbers of candidates ($k$) processed by GPT-4o-mini in \algzsgpt, comparing it against the use of LLaMA3.3-70B (Llama) and \algzsbp (BP).}

\revision{The LLM-based re-ranking approach improves over the bipartite. For example, at $k=5$, we observe a 6.7\% improvement in MRR over the bipartite baseline, rising from 0.731 to 0.780.
Recall@GT shows even more promising gains, from 0.375 to 0.475—a 26.7\% increase.
Increasing $k$ enhances MRR, although non-forwarded candidates score lower, reducing Recall@GT which considers all pairs. 

When all schemas sent to the LLM reranker, both  GPT-4o-mini (All Schema) and LLaMA3.3-70B (Llama), the performance is lower than all \alg variations. 
Furthermore, runtime increases with column count and is substantial when reranking all schemas, reaching over 6,000 seconds per table for GPT-4o due to LLM API latency. In contrast, when the SLM-based retriever is used, the runtime remains practical. We used $k=20$ for our other experiments, as it balances good MRR, Recall@GT, and runtime across \GDC and other datasets.
These results demonstrate the effectiveness of combining SLMs and LLMs for \smabrev.
}
\section{Related Work}
\label{sec:relatedwork}
This section discusses schema and ontology matching approaches related to our work.

\myparagraph{Traditional Methods}
 A straightforward approach for \smabrev is detecting overlap in column names~\cite{cupid2001,melnik2002similarity,coma2002} and overlap in column values~\cite{comaplusplus2005,attributediscovery2011}. Some incorporate relaxations when measuring overlaps, such as accounting for syntactic and semantic similarities between column names and values~\cite{cupid2001,comaplusplus2005}. Others also consider factors like data type relevance and value distribution~\cite{comaplusplus2005,attributediscovery2011}. Among these, the \coma algorithm stands out for integrating the most strategies and weighting their outputs to achieve better accuracy~\cite{coma2002, comaplusplus2005}, often remaining competitive even against more recent approaches~\cite{valentine2021}.
However, these approaches often struggle to capture complex relationships and deeper semantics within datasets~\cite{khatiwada2023santos}.

\myparagraph{Small Language Model-Driven Methods}
Methods based on small language models (SLMs) usually use embeddings to encode and compare column data~\cite{cappuzzo2020creating}.
Contrastive learning can improve an SLM's ability to distinguish matching and non-matching column pairs~\cite{starmie2023, cong2023pylon,schemaMatchPreTrainedICDE2023}, while synthetic tabular data generation can help models to improve without ground truth~\cite{starmie2023,inSituSchMatcICDE2024}.
Among these methods, \texttt{ISResMat} customizes pre-trained models for dataset-specific adaptation, generating training pairs from table fragments, and applying pairwise comparison losses to refine matching accuracy~\cite{inSituSchMatcICDE2024}. 
\texttt{Unicorn}, a general matching model, employs contrastive learning and a Mixture-of-Experts (MoE) layer within its architecture to discern matches but relies on supervised training~\cite{tu2023unicorn}.

\myparagraph{Large Language Model-Driven Methods}
Recent works have leveraged large language models (LLMs) for various aspects of tabular data management, predominantly focusing on single-table tasks~\cite{chorus@vldb2024, hegselmann2023tabllm}. Some studies discuss LLM applications for \smabrev  and highlight the potential of LLMs for this task~\cite{tableGPT2024,archetypeVLDB2024}. However, these approaches often rely solely on prompting strategies—either fine-tuned or zero-shot—which suffer from scalability issues and high computational costs\revision{~\cite{tableGPT2024, parciak2024schema, fang2024large, feng2024cost}}.
~\citet{sheetrit2024rematch} and~\citet{zhang2024smutf} utilize zero-shot pre-trained LLMs for \smabrev.~\citet{sheetrit2024rematch} addresses multiple-table matching, whereas our work focuses on two-table \smabrev, particularly for tables with numerous columns.
\citet{zhang2024smutf} uses rule-based feature extraction and trains an XGBoost classifier with gold data, a supervised approach distinct from our unsupervised method.
\revision{
\citet{xu2024kcmf} proposed manually-derived rules to guide LLMs during the matching process and incorporated external knowledge to deal with hallucinations. \citet{parciak2024schema} explored various prompting strategies for matching source attributes to a target schema. Note that the evaluation of these methods only take into account table/attribute names and descriptions~\cite{xu2024kcmf,parciak2024schema}---unlike \alg, they do not consider values.}  

\myparagraph{\revision{Ontology Matching}} \revision{
Ontology matching (OM) is related to schema matching (SM) but focuses on identifying semantic correspondences between elements (e.g., classes and properties) across different ontologies expressed in languages like OWL and RDF/XML~\cite{ontoMatchBook2013}.
Some OM approaches use heuristics, rule-based methods~\cite{ontoMatchBook2013, melt2019}, structural similarity, linguistics, and domain-specific resources~\cite{LogMap2012, logmapBio2017}. To facilitate the matching of domain-specific terminology, semi-automatic annotation using vocabularies and ontologies has been used to enrich schema labels~\cite{norms2011, BENEVENTANO2015119}.
Recent OM approaches leverage LLMs and, like \alg, use a two-phase matching process~\cite{LLMs4OM2025, olala2023}.
OLaLa uses embeddings for candidate matches and LLM evaluation with natural language conversion~\cite{olala2023}, while LLMs4OM employs RAG to extract and classify concept similarities~\cite{LLMs4OM2025}. Unlike these methods that prompt per candidate pair, \alg efficiently generates a ranked list for each input column.
}




\revision{
Adapting OM methods to SM is not straightforward: OM approaches target structured relationships rather than flat tabular data, may disregard the similarity of column values, and require format conversions that can impact performance.
We compared \alg against two OM systems, LogMap~\cite{LogMap2012} and LLMs4OM~\cite{LLMs4OM2025}, by treating tables as classes and columns as properties.
They underperformed SM-based baselines by a large margin.
Nonetheless, given their shared goals and challenges, combining OM and SM techniques is an interesting direction for future work. 
}


\section{Conclusions and Future work} \label{sec:conclusion}
We proposed \alg, a framework that leverages small and large language models to derive schema matching strategies that generalize across domains and balance accuracy and runtime tradeoffs. We introduced a new benchmark that captures some of the complexities in biomedical data integration and presents new challenges for schema matching. With a detailed experimental evaluation, including comparisons against state-of-the-art methods and ablations, we demonstrate the effectiveness of \alg and our design choices.

\revision{
There are several directions for our future work.
 \alg’s accuracy depends on SLM retrieval—if the correct match is missing from the top-$k$, reranking cannot recover it. We plan to explore hybrid retrieval strategies and training SLMs on large, multi-domain datasets. To address limitations in underrepresented domains, we also aim to integrate external knowledge for better zero-shot performance.
Reranking quality can vary due to prompt dependence—a known limitation of instruction-tuned LLMs~\citep{archetypeVLDB2024}. Future work may leverage prompt tuning frameworks like DSPy~\citep{khattab2024dspy} for systematic, task-aware optimization.
While LLM-based reranking improves accuracy, it introduces cost overheads. Future work could optimize reranking with smaller, efficient LLMs.
}

\section*{Acknowledgments}
This work was supported by NSF awards IIS-2106888 and OAC-2411221,  the DARPA
ASKEM program
Agreement No. HR0011262087, and the ARPA-H BDF program. The views, opinions, and findings expressed are those of the authors and should not be interpreted as representing the official views or policies of the DARPA, ARPA-H, the U.S. Government, or NSF. We thank our collaborators from the NYU School of Medicine, David Fenyo, Wenke Liu, and Sarah Keegan, for sharing their experience in biomedical data integration and contributing to the GDC benchmark. We also thank Roque Lopez for his contributions to the code.

\balance

\bibliographystyle{ACM-Reference-Format}
\bibliography{biblio}

\end{document}